\documentclass[aps,prmaterials,reprint,superscriptaddress]{revtex4-2}

\usepackage{amsmath}
\usepackage{dcolumn}
\usepackage{bm}
\usepackage{amssymb}
\usepackage{color}
\usepackage{tabularx}
\usepackage{graphicx}
\usepackage[normalem]{ulem}
\usepackage{soul}

\newcommand \be {\begin{equation}}
\newcommand \ee {\end{equation}}
\newcommand \ba {\begin{eqnarray}}
\newcommand \ea {\end{eqnarray}}

\begin{document}
\title{Uncovering liquid-substrate fluctuation effects on crystal growth and disordered hyperuniformity of two-dimensional materials}
\author{S. K. Mkhonta}
\affiliation{Department of Physics, University of Eswatini, Private Bag 4,
Kwaluseni, M200, Eswatini}
\affiliation{Department of Physics, Oakland University, Rochester, Michigan 48309, USA}
\author{Zhi-Feng Huang}
\affiliation{Department of Physics and Astronomy, Wayne State University,
Detroit, Michigan 48201, USA}
\author{K. R. Elder}
\affiliation{Department of Physics, Oakland University, Rochester, Michigan 48309, USA}
\date{\today}

\begin{abstract}
 We investigate the growth of two-dimensional (2D) crystals on fluctuating
 surfaces using a phase field crystal model that is relevant on atomic 
 length and diffusive time scales. Motivated by recent experiments which
 achieved unprecedented fast growth of large-size high-quality 2D crystals on
 liquid substrates, we uncover novel effects of liquid surfaces 
 on microstructural ordering. We find that substrate fluctuations generate
 short-ranged noise that speeds up crystallization and grain growth of the
 overlayer, surpassing that of free-standing system.  Coupling to the 
 liquid substrate fluctuations can also modulate local randomness, 
 leading to intriguing disordered structures with hidden spatial order, i.e., 
 disordered hyperuniformity. These results reveal the physical mechanisms 
 underlying the fast growth of 2D crystals on liquid surfaces and 
 demonstrate a novel strategy for synthesizing disordered hyperuniform 
 thin film structures. 
 \end{abstract} 
 \maketitle

\section{Introduction}
An overarching mission in materials research is to design robust strategies to
enable the assembly of billions of atoms, molecules, or tiny building blocks
into the desired high-quality macroscopic structures. A recent promising 
technique for rapidly growing large-scale, defect-free two-dimensional (2D) 
crystals such as graphene, hexagonal boron nitride (hBN), and transition metal
dichalcogenides (TMDs) is via chemical vapor deposition on substrates of liquid
metals~\cite{Geng2012,Jankowski2021,Lee2018,Zheng2019} 
or other molten systems~\cite{Chen2017}.

The growth of these 2D crystals can occur 
at high temperatures over 1000$^{\circ}$C, and
thus a broad family of possible substrates such as 
liquid-state metals (Cu, In, Ni, Ga, and Sn) and molten 
glasses~\cite{Zheng2019,Chen2017,Zeng2014}
can be used.  Molten solids appear as an atomically isotropic fluid, 
and are smooth (without kinks, terraces or protrusions) as compared to
their solid-state counterpart. On ultra-flat liquid surfaces crystallites can nucleate anywhere, move and rotate freely~\cite{Zeng2018}. These factors contribute to the emergence of
high-quality large crystal flakes, e.g., 100 $\mu$m in 
graphene~\cite{Geng2012,Jankowski2021} or hBN~\cite{Lee2018} 2D sheets within a few minutes at remarkable growth rates which are 2 to 6 orders of magnitude higher than that via solid substrates \cite{Geng2012,Lee2018}.

The mechanisms underlying the fast growth rate of large-size 2D crystals on
liquid-state surfaces are still not well understood. 
Recent studies have been focused on real-time \textit{in situ}
experiments \cite{Jankowski2021} and atomistic simulations including 
kinetic Monte Carlo (kMC) and Molecular Dynamics (MD) simulations \cite{Xu2020,Burov2022}. 
Here we develop a density-field model based on the phase field crystal (PFC)
framework \cite{Elder2002,Elder2004,Elder2007,Greenwood2010,Mkhonta2013} 
which has been successful in describing microstructure evolution at diffusive 
time scales and atomic length scales that are currently inaccessible to 
kMC and MD type atomistic simulations.  The PFC method is a powerful and 
computationally efficient tool for studying epitaxial growth, such as 
nanostructure formation and growth during strained film epitaxy 
\cite{Huang2008,Huang2010,Wu2009} and the characterization of commensurate 
to incommensurate transition in heteroepitaxial thin films and 
overlayers \cite{Elder2012,Elder2016}.
The growth of a 2D crystal on a solid amorphous (glass)
substrate has also been explored using the PFC method~\cite{Granato2011}, 
which revealed the amorphization of the adlayer, consistent with the effects 
of quenched disorder on crystalline structures. 

In this work we consider a critical factor of solidification on liquid substrate, 
the random noise generated by liquid-substrate density fluctuations. The coupling between the overlayer atomic density and the liquid-substrate fluctuations changes the system dynamics and equilibrium properties in a fundamental manner.
 These include not only the fast dynamics of 2D growth,
but also the emergence of disordered hyperuniformity, which is an exotic disordered 
state with a hidden long-range order for which the system structure factor 
scales as $S(q) \sim q^\beta$ as $q \rightarrow 0$ (with exponent $\beta >0$) 
\cite{Torquato2003}, showing the suppression of long-wavelength fluctuations. 
This type of hyperuniformity has been observed in experiments of phase-separated 
DNA droplets~\cite{Wilken2023} and spherical block copolymer micelles~\cite{Zito2015} 
and in simulations of phase separating star or brush-like-polymers
\cite{Chremos2018,Chremos2020}. Disordered hyperuniform materials lie
between the extreme of order and disorder and are desirable in the development of novel 
material systems with e.g., large isotropic photonic band gap~\cite{Yu2021,Man2013}, 
superior optical absorption in photovoltaics~\cite{Oskooi2012,Pala2013}, and 
high temperature superconductivity~\cite{Thien2017,Llorens2020}.

\section{The PFC model}

We consider the PFC dynamical equation of the system in terms of the rescaled adlayer particle density variation field $\psi(\mathbf{x},t)$, i.e., 
\be 
\frac{\partial}{\partial t}\psi(\mathbf{x},t)
= \nabla^2\mu(\mathbf{x},t) + \bm{\nabla}\cdot\bm{\eta}(\mathbf{x},t), 
\label{eq.dynamics} 
\ee 
where $\mu$ is the system chemical potential and $\bm{\eta}(\mathbf{x},t)$ is a 
Gaussian white noise field with zero mean and variance $\langle \eta(\mathbf{x},t) \eta(\mathbf{x'},t')\rangle
=2D\delta(\mathbf{x}-\mathbf{x'})\delta(t-t')$, describing thermal fluctuations of the adlayer particles with $D \propto T$ according to the fluctuation-dissipation relation. 
The chemical potential $\mu$ is expressed as the functional derivative of the free energy functional $\mathcal{F}$, i.e., $\mu =\delta \mathcal{F}/\delta \psi$. For 2D growth on a fluctuating surface, we set
\be 
\mathcal{F}= \int d\mathbf{x} \left\{
\frac{\psi}{2} \left [ r+ \left ( 1 +\nabla^2 \right )^2 \right ] \psi
+\frac{\psi^4}{4} + f\psi \right\}, 
\label{eq.energy} 
\ee 
where the first two terms correspond to those of the traditional PFC model free energy \cite{Elder2002,Elder2004,Elder2007}, and the last term is added phenomenologically to describes the coupling between the adlayer particle density and the liquid-substrate density fluctuations that are represented by a fluctuating random field $f(\mathbf{x},t)$. This form of coupling is motivated by a derivation based on the classical dynamic density functional theory (DDFT) for a PFC free energy of binary systems (see Appendix A for details). For a given adlayer average density $\psi_0$, in the absence of substrate coupling the 2D system free energy is minimized by the
homogeneous (liquid), stripe, and triangular ($\psi_0 <0$) or honeycomb ($\psi_0>0$) phases,
depending on the temperature control parameter $r\propto (T-T_m)/T_m$, with temperature $T$ and the melting point $T_m$.

Here we consider the substrate noise $f(\mathbf{x},t)$ to have short-ranged correlations
\be
\langle f(\mathbf{x},t) f(\mathbf{x'},t')\rangle
=2V\delta(\mathbf{x}-\mathbf{x'})\delta(t-t'),
\label{eq.randomF}
\ee
where $V$ is the adlayer-substrate coupling strength that is dependent on microscopic details of the interaction (as demonstrated in Appendix A).  In reality, $f(\mathbf{x},t)$  can be described by a longer range
two-point correlation function. However, the current approximation,
Eq.~(\ref{eq.randomF}), well captures the large lengthscale physics of the
problem since the liquid structure factor is finite at small wave vector
$\mathbf{q} \rightarrow 0$. The delta-function approximation for a random
potential has been shown to correctly capture the long distance physics in other
periodic systems~\cite{Toner1997,Toner1999, Bellini2001} and produce
consistent results with other quasi long-range disorder potential within the PFC
model~\cite{Granato2011}.
In addition, a recent study showed that Gaussian white noises are relevant for the emergence of hyperuniformity in reaction-diffusion systems of active and passive particles \cite{Ma2023}.

Hence the system dynamics in Eq.~(\ref{eq.dynamics}) contains two types 
of conserved noise: the substrate-induced fluctuations, $\nabla^2 f$, and the
intrinsic noise of overlayer, $\bm{\nabla} \cdot \bm{\eta}$. The former is greatly suppressed
at long distances due to the Laplacian operator. Thus, these two noise fields
contribute differently to large scale effects, as detailed below.

\section{Results}

\subsection{Analytical results}

Linearizing Eq.~(\ref{eq.dynamics}) about the homogeneous state with $\psi(\mathbf{x},t)=\psi_0$ gives the following equation in Fourier space
\be
\frac{d}{dt}\hat{\psi}_\mathbf{q} =-q^2G(q)\hat{\psi}_\mathbf{q} -
q^2\hat{f}_\mathbf{q} +i\mathbf{q}\cdot \hat{\bm \eta}_\mathbf{q},
\label{eq.motion2} 
\ee 
where $G(q) = r+3\psi_0^2+ (1-q^2)^2$ and $\hat{\psi}_\mathbf{q}$, 
$\hat{f}_\mathbf{q}$, and $\hat{\bm \eta}_\mathbf{q}$ are the Fourier transforms of $\psi$, $f$, and $\bm{\eta}$.  
Equation~(\ref{eq.motion2}) can be solved exactly leading to a prediction 
for the structure factor, 
$S({\mathbf{q}},t) = \langle |\hat{\psi}_{\mathbf{q}}|^2 \rangle$, i.e.,  
\be
S(\mathbf{q},t) = \frac{D + Vq^2}{G}  \left(1-e^{-2q^2Gt} \right).
\label{eq.SF}
\ee 
At high temperatures $G(q)>0$ for all $q$, and the steady state solution of
Eq.~(\ref{eq.SF}) is given by
\be 
S^{\rm eq}(q)= \frac{D +Vq^2}{|G|}.
\label{eq:Slin}
\ee 
In the small-$q$ limit,
\be
S^{\rm eq}(q\rightarrow 0) = cD+ c(V +2cD)q^2+\cdots, 
\ee 
where $c=1/(r+3\psi_0^2+1)$. When $D$ is vanishingly small, 
$S^{\rm eq}_q \propto q^2$ and the system is hyperuniform, indicating that
the steady state fluctuations are greatly suppressed at small $q$. 
Interestingly, Eq.~(\ref{eq.motion2}) reduces to the generalized Langevin 
equation for systems where hyperuniformity is a result of competition 
between diffusion and short-range noise \cite{Hexner2017}, while here
it is the liquid-substrate noise that causes the $q^2$ scaling at small $q$.

At temperatures below the adlayer melting temperature, $r+3\psi_0^2<0$
and linear instability occurs.  At the onset of this
instability $S({\mathbf{q}},t)$ is very small, such that Eq.~(\ref{eq.SF}) can 
be applied to obtain the approximation 
\be
S(\mathbf{q},t) \approx 2q^2(D+q^2V)t + \cdots.
\label{eq.SFDyn}
\ee
This implies the liquid-substrate fluctuations contribute to rapid
crystallization and grain growth at the early stage of overlayer ordering,
with faster crystallization rate for system of large $V$. 
From Eq.~(\ref{eq.SFDyn}) we can derive a length scale $l_h =\sqrt{V/D}$ 
below which the substrate fluctuation effects dominate the dynamics. 
At later times nonlinear effects play a key role and
the above linear theory breaks down. 

\begin{figure*}[hbt]
\centerline{\includegraphics[width=\textwidth]{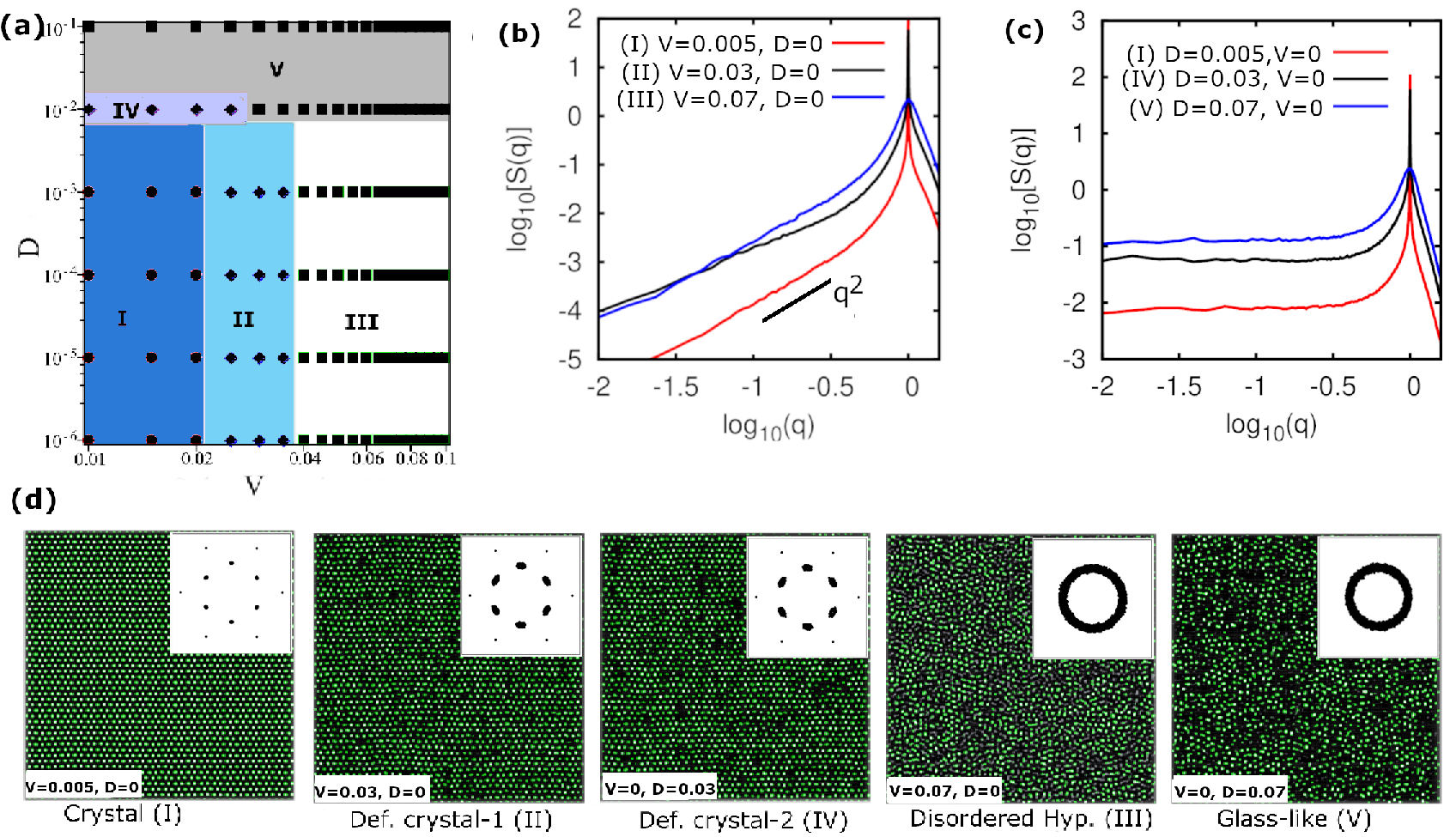}}
\caption{(a) Structural phase diagram for monolayers on a liquid substrate,
including three hyperuniform states of (I) prefect crystal, (II)
defective crystal-1, and (III) disordered hyperuniform phase, and two
non-hyperuniform states of (IV) defective crystal-2 and (V) disordered state. 
The dots correspond to simulation results. 
In (b) and (c) the circularly averaged structure factor is shown for different structures. 
(d) Sample simulation snapshots of the density profile (for $1/16$ of the simulation box), 
where light-green areas correspond to particle density peaks. 
The corresponding diffraction patterns are shown as insets.
} 
\label{figure1} 
\end{figure*}

\subsection{Numerical results}

\subsubsection{Structural phase diagram}

To study nonlinear effects, Eq.~(\ref{eq.dynamics}) was solved numerically on 
a 2D grid of $1024\times 1024$ points with a mesh size $\Delta x =\pi/4$. A pseudo-spectral algorithm~\cite{Granato2011,Warren2008} is used with periodic
boundary conditions and a time step $\Delta t =0.5$. We set $\psi_0=r=-0.28$, 
for which the deterministic equilibrium phase is triangular~\cite{Elder2004}. To determine a structural phase diagram in the $(D,V)$ plane, we performed simulations for each point in the parameter space numerous times with the initial density profile being the deterministic one-mode triangular solution (given in Ref.~\cite{Elder2004}). Each simulated system was allowed to relax for a sufficiently long period up to $t=10^5\Delta t$. The final-state structures were distinguished in terms of the hyperuniformity, 2D structure factor, and density patterns, with results shown in Fig.~\ref{figure1}. We ascertained that random initial conditions reached the same final states (albeit in longer times or for smaller regions).

In the small-$D$ limit, the system exhibits three hyperuniform states  as $V$
increases, namely a perfect crystal, a defective crystal-1, and a disordered
hyperuniform state. It is interesting to note that the scaling of the circularly
averaged structure factor $S(q)$ for defective crystal-1 deviates from $S(q)
\sim q^2$, giving $S(q) \sim q^{\beta}$ with $1 \leq \beta \leq 2$. This state consists of hyperuniformity-preserving topological defects (as also observed in amorphous 2D networks \cite{Chen2020}), as induced by local perturbations which do not significantly change density fluctuations of the parent lattice at large lengthscales.

At large $D$, the system exhibits two
non-hyperuniform states: a defective crystal-2 and a disordered state. 
To identify hyperuniformity, we utilized the normalized structure factor 
\be 
\bar{S}(q) =S(q)/S(q_{\rm peak}), 
\ee
where $q_{\rm peak}$ is the wave number at the first peak of the structure factor.
We classified systems as being hyperuniform if $\bar{S}(q\rightarrow 0) \lesssim 10^{-3}$ \cite{Torquato2016}. Hence defective crystal-2 is
non-hyperuniform as seen from Fig.~\ref{figure1}(c), while a perfect crystal state still exhibits hyperuniformity at small enough nonzero $D$ of overlayer white noise even without the substrate effect, although with different scaling of $S(q)$ as compared to that dominated by liquid-substrate fluctuations [see Fig.~\ref{figure1}(c) vs \ref{figure1}(b)].
Generally, at large enough
fluctuations (with large $V$ or $D$) crystalline order disappears, leading to the 
formation of disordered states which could be hyperuniform when $l_h$ is of the 
order of or larger than system size.

\begin{figure}[htb]
\centerline{\includegraphics[width=0.45\textwidth]{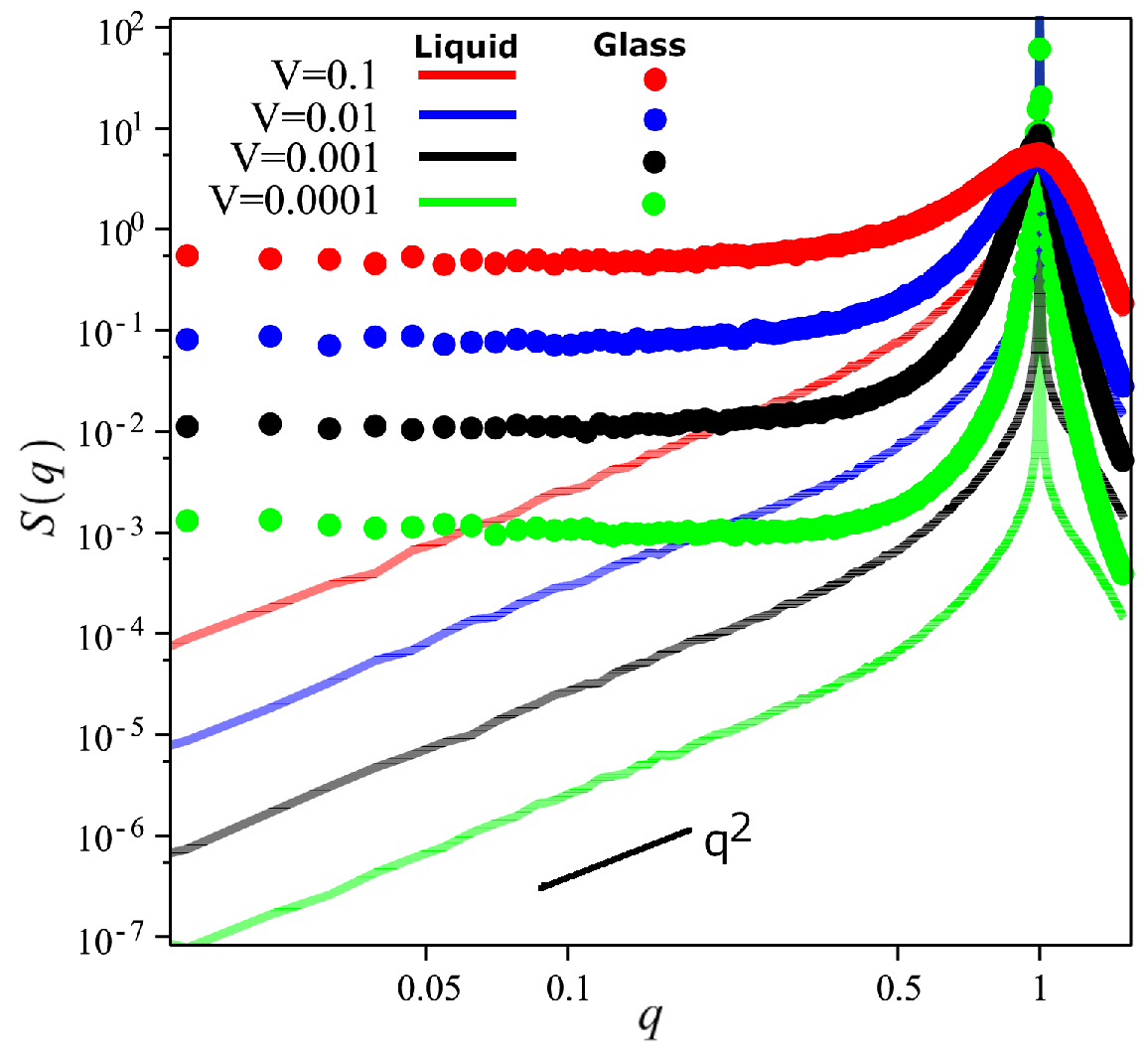}}
\caption{Systems structure factor at different liquid and glass substrate coupling strength with $D=0$.} 
\label{figure2} 
\end{figure}

\subsubsection{Difference between liquid and glass substrates}

We also compare ordering on a liquid substrate
against that on a solid glass-like substrate. The later can be represented by a
quenched disordered field, $f(\mathbf{x})$, which is random 
and short-range correlated in space and time-independent. We used the same
PFC dynamic equation (\ref{eq.dynamics}) while enforcing that $\langle f(\mathbf{x})
f(\mathbf{x'})\rangle =2V\delta(\mathbf{x}-\mathbf{x'})$ for glass substrate,
as done in previous studies~\cite{Granato2011} which
showed that a quenched random field can destabilize crystallization 
in a large-sized system even in the zero-$D$ limit.
This has been confirmed in our simulations. In contrast, this effect of disordering does not occur for a liquid substrate at modest $V$ since the substrate
pinning sites are then dynamic. As shown in Fig.~\ref{figure2}, at 
small and modest $V$ and $D=0$
while quenched disorder on a glass substrate leads to non-hyperuniform disordered 
state, the structures forming on the liquid substrate always remain
hyperuniform.  

\begin{figure*}[htb]
\centerline{\includegraphics[width=\textwidth]{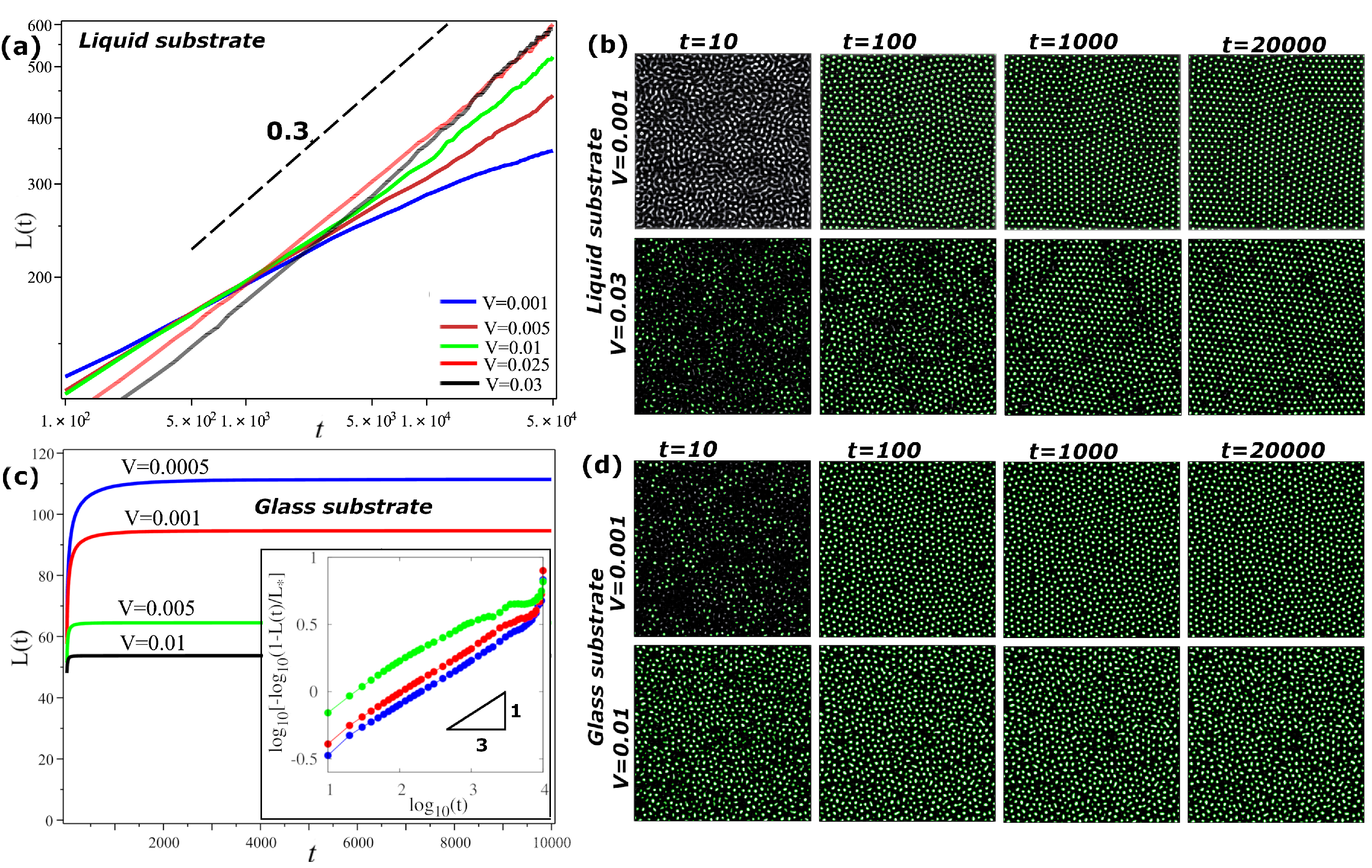}}
\caption{Dynamical behavior
at various coupling strengths ($V$) at $D=0$. Results in (a) and (c) have been 
averaged over 50 simulation runs. The inset in (c) is to demonstrate that the
domain size $L(t)$ grows consistently with the stretched exponential function
(with a scaling exponent $\alpha= 1/3$). 
Snapshots in (b) and (d) correspond to $1/16$ of the simulation box.} 
\label{figure3}
\end{figure*}

The contrast between the two substrates can be seen during the grain growth 
from a quenched liquid at $D=0$, for which the initial conditions were set as 
$\psi(\mathbf{x},0)= \psi_0$. 
Figure~\ref{figure3} shows some snapshots of the simulated structures that
form on two types of substrates. To characterize the growth process,
the dominant peak of
$S(q,t)$ was fit to a squared Lorentzian \cite{Cross1995,Hou1997}.
The inverse of the full width at half maximum of the peak was then used
as a measure of the average grain size, $L(t)$. As seen in Fig.~\ref{figure3}(a),
higher coupling strength $V$ leads to more rapid grain growth on the liquid 
substrate, due to a stronger influence of substrate fluctuations on assisting the 
ordering of the monolayer to rapidly escape any metastable configurations.
The grain size follows a power-law scaling of $L(t)\sim t^n$, with 
$n\rightarrow 1/3$ as $V$ increases. These results are consistent with previous
simulations of noise-driven domain growth in free-standing modulated
monolayers, such as stripes ~\cite{Hou1997,Boyer2002} and hexagonal
patterns~\cite{Shiwa2011}. 

However, the influence of a glass substrate is fundamentally different.
As shown in Fig.~\ref{figure3}(d), crystallization is hindered on the glass 
substrate at strong enough couplings as a result of substrate pinning.  
Here $L(t)$ rapidly increases
with time before reaching a plateau, but with sizes much smaller than
those for liquid substrate. The domain growth and stagnation are well fit to 
the classical Johnson-Mehl-Avrami-Kolmogorov equation for polycrstalline solidification over three decades in time 
[see the inset of Fig.~\ref{figure3}(c)], i.e., a stretched exponential behavior
\begin{equation}
L(t) =L_{\infty} [1-\exp(-kt^\alpha)],
\end{equation}
with $\alpha =0.33\pm 0.01$ which is independent of $V$.  Interestingly, 
this value of scaling exponent $\alpha$ is consistent with that
observed in the grain growth of doped colloidal polycrystals~\cite{Hutchinson2022}.
The other parameters, including the characteristic pinning lengthscale, $L_{\infty}$, and the domain size growth 
rate, $k$, both depend on the substrate coupling strength $V$, as shown in Table \ref{Table1}.

\begin{table}
\caption{\label{Table1} The dependence of characteristic pinning lengthscale, $L_{\infty}$, and the domain size growth 
rate, $k$, on the liquid-substrate coupling strength $V$.}
\begin{ruledtabular}
 \begin{tabular}{ccc}
   $V$ & $L_{\infty}$ & $k$\\
 \hline
   0.0005 & $111.40\pm 0.01$ & $0.400\pm 0.002$ \\
   0.001 & $94.60\pm 0.08$ & $0.484\pm 0.003$\\
   0.005 & $64.50\pm 0.01$ & $0.803\pm 0.006$ \\
 \end{tabular}
 \end{ruledtabular}
 \end{table}
 
\begin{figure*}
\centerline{\includegraphics[width=\textwidth]{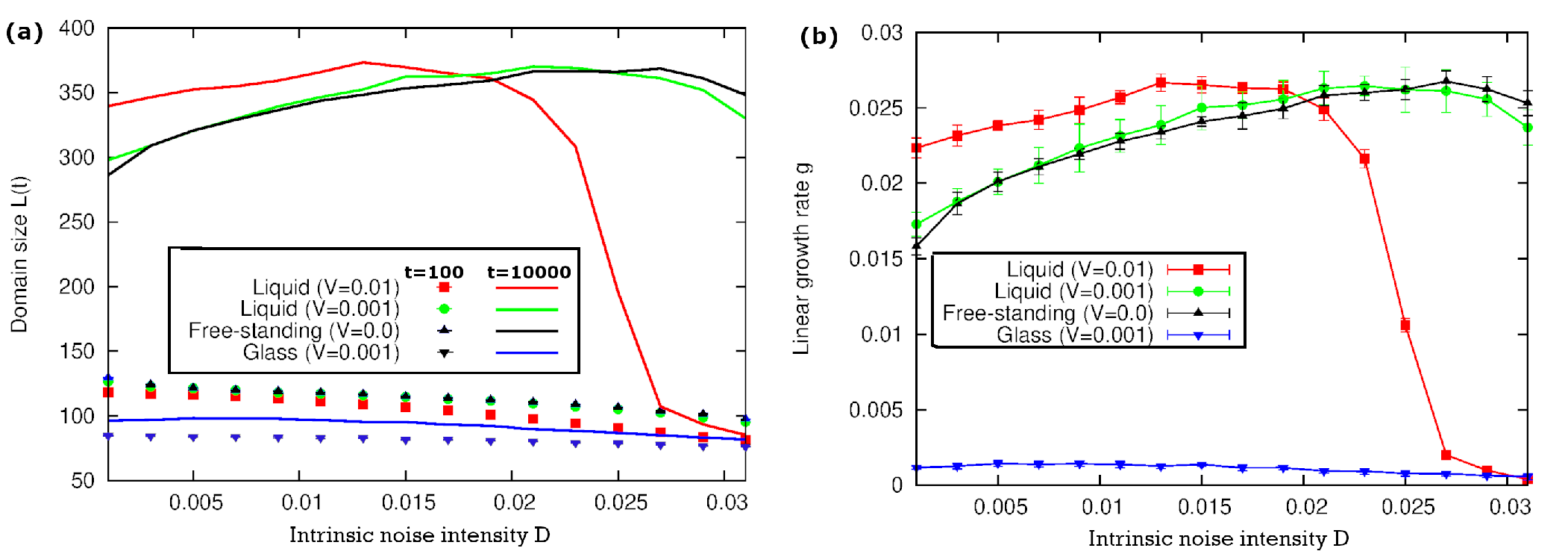}}
\caption{(a) Dependence of domain size $L$ on the intrinsic noise strength $D$. 
(b) Average growth rate over the time interval $t\in [100,10000]$. Each point 
is the statistical average over 50 independent runs, with initial condition $\psi(\mathbf{x},0)=\psi_0$.} 
\label{figure4} 
\end{figure*}

\subsubsection{Fast growth on a liquid substrate}

Figure~\ref{figure4} shows the effect of intrinsic noise strength, $D$,
under three conditions of the
 monolayer: free-standing ($V=0$), on a liquid, 
and on a glass substrate.   
At the early stage ($t=100$), the average domain size
of the emerging polycrystals decreases as $D$ increases,  which can be attributed to faster solidification process at larger noise
intensities \cite{Laszlo2011,Laszlo2011b}.
Generally, the free-standing case ($V=0$) is characterized by larger grains at
the early stage since it has lowest solidification rate [see
Eq.~(\ref{eq.SFDyn})].  Conversely, at the late time stage ($t=10000$) larger domains appear in noisier systems. However, there is a transition to disordered state for large enough $D$. The transition occurs at largest $D$ for the free-standing case since it contains only one noise source.  

To further understand these results we calculated the
mean linear domain growth rate  $g(t) =\Delta L/\Delta t$ over
time interval $t \in [100, 10000]$, as shown in Fig.~\ref{figure4}(b).
Interestingly, the liquid substrate at large $V$ ($=0.01$) produces the largest
growth rates over a wide range of $D$, exceeding that of free-standing case
(which is inaccessible experimentally).  Again, this emphasizes the role of
fluctuations in fast-tracking the crystallization process, with more noise
sources leading to faster rates of pattern formation. 
It is noted that the results here correspond to the
late stage of solidification in experimental systems where the dynamics are
associated with defect annihilation and grain boundary motion in a
polycrystalline sample. The experiments of graphene or hBN growth
\cite{Geng2012,Jankowski2021,Lee2018,Zheng2019} have explored the nucleation and
growth of crystallites before merging to form large crystals or polycrystals,
which could be examined in the solid-liquid coexistence regime of the PFC model
\cite{Elder2004,Laszlo2011,Laszlo2011b} (at a much smaller $|r|$, i.e., higher
growth temperature than the regime studied here).

\section{Summary and Outlook}

In summary, we have modeled and simulated the growth dynamics of 2D monolayers 
on a fluctuating surface. Our results reveal that time-varying liquid
substrate fluctuations can generate disordered hyperuniform structures and also
significantly enhance the growth dynamics.  We, thereby, provide a physical
mechanism that contributes to the superfast growth of large-size 2D crystals
observed in recent experiments.  Our theory is applicable to a variety of modulated
or microphase separating systems that can be represented by a PFC-type free energy  functional and dynamics, such as disordered hyperuniformity in star-shaped block  copolymers and micelles, nanosheets of block copolymers \cite{Vargo2023},  and colloidal crystals in liquid-liquid or liquid-air 
interfaces \cite{Bruevich2019,Vialetto2021,Ke2023}. 

A future in-depth study will utilize material-specific parameters for 2D systems such as graphene or hBN grown on liquid substrates, to predict and optimize growth conditions of these systems. This can be achieved via matching the PFC model parameters to lattice symmetry, elastic modulus, and lattice spacing for these materials as conducted in previous studies \cite{Hirvonen2016,Taha2017}. A key factor here is the value of the liquid-film coupling strength which is currently unknown and difficult to identify, and requires further atomistic calculations (via e.g., MD).  In the context of polycrystal ordering, previous studies \cite{Granato2011} have shown that monolayer movements even on disordered solid substrate can enhance ordering, and grain rotation has also been examined in PFC simulations of 2D materials including graphene and hBN \cite{Waters2022}. Therefore, the interplay of the substrate fluctuations and possible movements of the overlayer crystallites or grains should further enhance the crystallization process, which can be quantified through this PFC modeling. In addition, quantitative analysis of fast-growth of 2D crystals on liquid substrates can address the effects of capillary fluctuations \cite{Harutyunyan2012} on e.g., single-crystal size limit and the hyperuniformity predicted here, which will require a PFC modeling incorporating long-wavelength out-of-plane deformations \cite{Granato2022}. 

\begin{acknowledgments}
S.K.M. and K.R.E. would like to acknowledge support from the National 
Science Foundation (NSF) under Grant No. DMR-2006456. Z.-F.H. acknowledges 
support from NSF under Grant No. DMR-2006446.
\end{acknowledgments}

\appendix
\section{Model derivation: Lowest-order approximation of the liquid-substrate coupling}

The system of an adlayer grown on top of a liquid substrate can be characterized as involving two sub-systems: sub-system $A$ consisting of adlayer atoms, and sub-system $B$ consisting of substrate surface atoms remaining at or close to thermodynamic equilibrium.  Within the PFC model framework, a binary $AB$ system can be described via two dimensionless coupled density fields, $n_A$ and $n_B$, representing the atomic number density variations of $A$ and $B$ sub-systems respectively. Based on the derivation from DDFT as shown in Refs.~\cite{Huang2010,Taha2019}, the rescaled free energy functional of this system is given by \cite{Taha2019,Taha2017}
\ba
\mathcal{F}_{a} &=&\int d\mathbf{x} \left[ -\frac{1}{2}\epsilon_A n_A^2 +\frac{1}{2}n_A(\nabla^2 + 1)^2n_A - \frac{1}{3}g_An_A^3\right.\nonumber\\
&& +\frac{1}{4}n_A^4+\frac{1}{2}\epsilon_Bn^2_B +\frac{1}{2}\beta_Bn_B(\nabla^2 +q_B^2)^2n_B \nonumber\\
&& -\frac{1}{3}g_Bn_B^3 +\frac{1}{4}vn_B^4 +\beta_{AB}n_A(\nabla^2+q_{AB}^2)^2n_B \nonumber\\
&&\left. +\alpha_{AB}n_An_B+\frac{1}{2}wn_A^2n_B +\frac{1}{2}un_An_B^2\right],
\label{Eq.FreeEnergyApp}
\ea
where $q_B =R_{AA}/R_{BB}$ and $q_{AB}=R_{AA}/R_{AB}$, with $R_{ij}$ ($i,j = A, B$) being the characteristic interparticle spacing for the corresponding atomic species. The rest of dimensionless parameters in Eq.~(\ref{Eq.FreeEnergyApp}) can be derived from  DDFT and expressed in terms of the Fourier components of two- and three-point direct correlation functions \cite{Taha2019}. 

Unlike the case of sublattice ordering considered in Refs.~\cite{Taha2017,Taha2019}, in this study subsystem $B$ (the liquid substrate) is assumed to be at or near thermodynamic equilibrium, i.e., $\epsilon_B >0$. Thus, the liquid-substrate homogeneous solution for $n_B$ can be written as
\be
n_B = n_B^0+ \delta n_B,
\label{Eq.LiqSub}
\ee
where $n_B^0$ is a constant representing the average substrate density, and $\delta n_B(\mathbf{x},t)$ describes deviations from the average density and fluctuates around the equilibrium state of the liquid substrate. Without loss of generality, we can set $n_B^0 =0$. We then substitute Eq.~(\ref{Eq.LiqSub}) into Eq.~(\ref{Eq.FreeEnergyApp}), retain quadratic terms involving $\delta n_B$, and consider the weak coupling between the film and liquid substrate (i.e., weak $A$-$B$ interaction). The $\alpha_{AB}$ and $\beta_{AB}$ coupling terms play essentially the same role particularly in the long wavelength limit, and thus only the $\alpha_{AB}$ term is kept. The higher-order couplings of $w$ and $u$ terms do not change properties of the system in an essential manner for the homogeneous state of $n_B$ and thus can be neglected. Further simplification can be made by incorporating the effect of $g_A$ cubic term into the $n_A^4$ quartic term \cite{Elder2004}. Taking all these assumptions into account, Eq.~(\ref{Eq.FreeEnergyApp}) reduces to 
\ba
\mathcal{F}_{a} &=&\int d\mathbf{x} \left[ -\frac{1}{2}\epsilon_A n_A^2 +\frac{1}{2}n_A(\nabla^2 + 1)^2n_A 
+\frac{1}{4}n_A^4\right.\nonumber\\
&& \left.+\alpha_{AB}n_A\delta n_B+\frac{\epsilon_B}{2}\delta n^2_B +\frac{\beta_B}{2}\delta n_B(\nabla^2 +q_B^2)^2\delta n_B \right]. \nonumber\\
\label{Eq.ReducedFE}
\ea
After changing variables via 
$n_A \rightarrow \psi$, $-\epsilon_A \rightarrow r$, and $\alpha_{AB}\delta n_B \rightarrow f$, the first four terms in Eq.~(\ref{Eq.ReducedFE}) lead to Eq.~(\ref{eq.energy}) representing the free energy functional for the atomic species in the overlayer.

The nature of the substrate fluctuation can be identified from solving the dynamical equation governing the substrate density field
\be
\frac{\partial n_B}{\partial t} = M_B\nabla^2\frac{\delta \mathcal{F}_a}{\delta n_B}+ \bm{\nabla} \cdot \bm{\eta}_B,
\label{eq.Sub_Dyn}
\ee
where $M_B$ is the mobility and the second term is the stochastic contribution associated with thermal fluctuations. The specific form of $\bm{\eta}_B$ is chosen to ensure that in the infinite time limit 
Eq.~(\ref{eq.Sub_Dyn}) gives rise to the correct equilibrium fluctuation spectrum for 
$n_B$ as introduced by Cook \cite{Cook70,Nik2010}.
This is achieved by setting $\bm{\eta}_B$ to be a Gaussian white noise satisfying the conditions
\ba 
\langle \bm{\eta}_B \rangle &=& 0,\nonumber\\
\langle \bm{\eta}_B(\mathbf{x},t)\bm{\eta}_B(\mathbf{x}',t') \rangle &=& 2 k_BT M_B
\delta(\mathbf{x}-\mathbf{x}')\delta(t-t'),~~
\ea
where $k_B$ is the Boltzmann constant and $T$ is the system temperature.

Substituting Eq.~(\ref{Eq.LiqSub}) into Eq.~(\ref{eq.Sub_Dyn}) we obtain
\ba
\frac{\partial \delta n_B}{\partial t} &=& M_B\nabla^2 \left [\epsilon_B+\beta_B(\nabla^2+q_B^2)^2  \right ]\delta n_B\nonumber\\
&& +\alpha_{AB} n_A+ \bm{\nabla} \cdot \bm{\eta}_B.
\ea
If neglecting the influence of the overlayer density $n_A$ on the variation of the homogeneous liquid substrate underneath as it is of higher-order contribution, and assuming that surface fluctuations are determined by the substrate liquid, in Fourier space the equilibrium state of liquid substrate corresponds to 
\be
\langle |\delta \hat{n}_B(q)|^2 \rangle = \frac{2k_BT}{\epsilon_B+\beta_B(q^2-q_{B}^2)^2}.
\ee 
At long enough length scales (e.g., $q << q_B$ in the long wavelength limit for the liquid state), we have
\be
\langle |\delta \hat{n}_B(q)|^2 \rangle \approx 2k_BT/(\epsilon_B+\beta_Bq_B^4) =\text{const.},
\label{Eq.relation}
\ee
which indicates that the liquid-substrate fluctuation, $\delta n_B(\mathbf{x},t)$, can be treated as a white noise at length scales beyond the microscopic atomic spacing. This leads to Eq.~(\ref{eq.randomF}) with short-ranged correlation as utilized in this work, i.e.,
$\langle f(\mathbf{x},t) f(\mathbf{x'},t')\rangle
=2V\delta(\mathbf{x}-\mathbf{x'})\delta(t-t')$,
where $V=\alpha^2_{AB}k_BT/(\epsilon_B+\beta_Bq_B^4)$ due to Eq.~(\ref{Eq.relation}) and $f(\mathbf{x},t) = \alpha_{AB}\delta n_B(\mathbf{x},t)$, when neglecting the high-wavenumber contributions at scales smaller than the atomistic ones. Note that the use of this delta-function form for liquid-substrate fluctuations without the filtering of short lengthscale fluctuations does not change the long-distance physics behavior of the system \cite{Granato2011}.  This argument has been demonstrated to hold also in other periodic systems \cite{Toner1997,Toner1999, Bellini2001}.

\bibliography{FinalRef}

\begin{thebibliography}{58}%
\makeatletter
\providecommand \@ifxundefined [1]{%
 \@ifx{#1\undefined}
}%
\providecommand \@ifnum [1]{%
 \ifnum #1\expandafter \@firstoftwo
 \else \expandafter \@secondoftwo
 \fi
}%
\providecommand \@ifx [1]{%
 \ifx #1\expandafter \@firstoftwo
 \else \expandafter \@secondoftwo
 \fi
}%
\providecommand \natexlab [1]{#1}%
\providecommand \enquote  [1]{``#1''}%
\providecommand \bibnamefont  [1]{#1}%
\providecommand \bibfnamefont [1]{#1}%
\providecommand \citenamefont [1]{#1}%
\providecommand \href@noop [0]{\@secondoftwo}%
\providecommand \href [0]{\begingroup \@sanitize@url \@href}%
\providecommand \@href[1]{\@@startlink{#1}\@@href}%
\providecommand \@@href[1]{\endgroup#1\@@endlink}%
\providecommand \@sanitize@url [0]{\catcode `\\12\catcode `\$12\catcode
  `\&12\catcode `\#12\catcode `\^12\catcode `\_12\catcode `\%12\relax}%
\providecommand \@@startlink[1]{}%
\providecommand \@@endlink[0]{}%
\providecommand \url  [0]{\begingroup\@sanitize@url \@url }%
\providecommand \@url [1]{\endgroup\@href {#1}{\urlprefix }}%
\providecommand \urlprefix  [0]{URL }%
\providecommand \Eprint [0]{\href }%
\providecommand \doibase [0]{https://doi.org/}%
\providecommand \selectlanguage [0]{\@gobble}%
\providecommand \bibinfo  [0]{\@secondoftwo}%
\providecommand \bibfield  [0]{\@secondoftwo}%
\providecommand \translation [1]{[#1]}%
\providecommand \BibitemOpen [0]{}%
\providecommand \bibitemStop [0]{}%
\providecommand \bibitemNoStop [0]{.\EOS\space}%
\providecommand \EOS [0]{\spacefactor3000\relax}%
\providecommand \BibitemShut  [1]{\csname bibitem#1\endcsname}%
\let\auto@bib@innerbib\@empty
%</preamble>
\bibitem [{\citenamefont {Geng}\ \emph {et~al.}(2012)\citenamefont {Geng},
  \citenamefont {Wu}, \citenamefont {Guo}, \citenamefont {Huang}, \citenamefont
  {Xue}, \citenamefont {Chen}, \citenamefont {Yu}, \citenamefont {Jiang},
  \citenamefont {Hu},\ and\ \citenamefont {Liu}}]{Geng2012}%
  \BibitemOpen
  \bibfield  {author} {\bibinfo {author} {\bibfnamefont {D.}~\bibnamefont
  {Geng}}, \bibinfo {author} {\bibfnamefont {B.}~\bibnamefont {Wu}}, \bibinfo
  {author} {\bibfnamefont {Y.}~\bibnamefont {Guo}}, \bibinfo {author}
  {\bibfnamefont {L.}~\bibnamefont {Huang}}, \bibinfo {author} {\bibfnamefont
  {Y.}~\bibnamefont {Xue}}, \bibinfo {author} {\bibfnamefont {J.}~\bibnamefont
  {Chen}}, \bibinfo {author} {\bibfnamefont {G.}~\bibnamefont {Yu}}, \bibinfo
  {author} {\bibfnamefont {L.}~\bibnamefont {Jiang}}, \bibinfo {author}
  {\bibfnamefont {W.}~\bibnamefont {Hu}},\ and\ \bibinfo {author}
  {\bibfnamefont {Y.}~\bibnamefont {Liu}},\ }\bibfield  {title} {\bibinfo
  {title} {Uniform hexagonal graphene flakes and films grown on liquid copper
  surface},\ }\href@noop {} {\bibfield  {journal} {\bibinfo  {journal} {Proc.
  Natl. Acad. Sci. U.S.A.}\ }\textbf {\bibinfo {volume} {109}},\ \bibinfo
  {pages} {7992} (\bibinfo {year} {2012})}\BibitemShut {NoStop}%
\bibitem [{\citenamefont {Jankowski}\ \emph {et~al.}(2021)\citenamefont
  {Jankowski}, \citenamefont {Saedi}, \citenamefont {La~Porta}, \citenamefont
  {Manikas}, \citenamefont {Tsakonas}, \citenamefont {Cingolani}, \citenamefont
  {Andersen}, \citenamefont {de~Voogd}, \citenamefont {van Baarle},
  \citenamefont {Reuter} \emph {et~al.}}]{Jankowski2021}%
  \BibitemOpen
  \bibfield  {author} {\bibinfo {author} {\bibfnamefont {M.}~\bibnamefont
  {Jankowski}}, \bibinfo {author} {\bibfnamefont {M.}~\bibnamefont {Saedi}},
  \bibinfo {author} {\bibfnamefont {F.}~\bibnamefont {La~Porta}}, \bibinfo
  {author} {\bibfnamefont {A.~C.}\ \bibnamefont {Manikas}}, \bibinfo {author}
  {\bibfnamefont {C.}~\bibnamefont {Tsakonas}}, \bibinfo {author}
  {\bibfnamefont {J.~S.}\ \bibnamefont {Cingolani}}, \bibinfo {author}
  {\bibfnamefont {M.}~\bibnamefont {Andersen}}, \bibinfo {author}
  {\bibfnamefont {M.}~\bibnamefont {de~Voogd}}, \bibinfo {author}
  {\bibfnamefont {G.~J.}\ \bibnamefont {van Baarle}}, \bibinfo {author}
  {\bibfnamefont {K.}~\bibnamefont {Reuter}}, \emph {et~al.},\ }\bibfield
  {title} {\bibinfo {title} {Real-time multiscale monitoring and tailoring of
  graphene growth on liquid copper},\ }\href@noop {} {\bibfield  {journal}
  {\bibinfo  {journal} {ACS Nano}\ }\textbf {\bibinfo {volume} {15}},\ \bibinfo
  {pages} {9638} (\bibinfo {year} {2021})}\BibitemShut {NoStop}%
\bibitem [{\citenamefont {Lee}\ \emph {et~al.}(2018)\citenamefont {Lee},
  \citenamefont {Choi}, \citenamefont {Yun}, \citenamefont {Kim}, \citenamefont
  {Boandoh}, \citenamefont {Park}, \citenamefont {Shin}, \citenamefont {Ko},
  \citenamefont {Lee}, \citenamefont {Kim} \emph {et~al.}}]{Lee2018}%
  \BibitemOpen
  \bibfield  {author} {\bibinfo {author} {\bibfnamefont {J.~S.}\ \bibnamefont
  {Lee}}, \bibinfo {author} {\bibfnamefont {S.~H.}\ \bibnamefont {Choi}},
  \bibinfo {author} {\bibfnamefont {S.~J.}\ \bibnamefont {Yun}}, \bibinfo
  {author} {\bibfnamefont {Y.~I.}\ \bibnamefont {Kim}}, \bibinfo {author}
  {\bibfnamefont {S.}~\bibnamefont {Boandoh}}, \bibinfo {author} {\bibfnamefont
  {J.-H.}\ \bibnamefont {Park}}, \bibinfo {author} {\bibfnamefont {B.~G.}\
  \bibnamefont {Shin}}, \bibinfo {author} {\bibfnamefont {H.}~\bibnamefont
  {Ko}}, \bibinfo {author} {\bibfnamefont {S.~H.}\ \bibnamefont {Lee}},
  \bibinfo {author} {\bibfnamefont {Y.-M.}\ \bibnamefont {Kim}}, \emph
  {et~al.},\ }\bibfield  {title} {\bibinfo {title} {Wafer-scale single-crystal
  hexagonal boron nitride film via self-collimated grain formation},\
  }\href@noop {} {\bibfield  {journal} {\bibinfo  {journal} {Science}\ }\textbf
  {\bibinfo {volume} {362}},\ \bibinfo {pages} {817} (\bibinfo {year}
  {2018})}\BibitemShut {NoStop}%
\bibitem [{\citenamefont {Zheng}\ \emph {et~al.}(2019)\citenamefont {Zheng},
  \citenamefont {Zeng}, \citenamefont {Cao}, \citenamefont {Zhang},
  \citenamefont {Gao}, \citenamefont {Xiao},\ and\ \citenamefont
  {Fu}}]{Zheng2019}%
  \BibitemOpen
  \bibfield  {author} {\bibinfo {author} {\bibfnamefont {S.}~\bibnamefont
  {Zheng}}, \bibinfo {author} {\bibfnamefont {M.}~\bibnamefont {Zeng}},
  \bibinfo {author} {\bibfnamefont {H.}~\bibnamefont {Cao}}, \bibinfo {author}
  {\bibfnamefont {T.}~\bibnamefont {Zhang}}, \bibinfo {author} {\bibfnamefont
  {X.}~\bibnamefont {Gao}}, \bibinfo {author} {\bibfnamefont {Y.}~\bibnamefont
  {Xiao}},\ and\ \bibinfo {author} {\bibfnamefont {L.}~\bibnamefont {Fu}},\
  }\bibfield  {title} {\bibinfo {title} {Insight into the rapid growth of
  graphene single crystals on liquid metal via chemical vapor deposition},\
  }\href@noop {} {\bibfield  {journal} {\bibinfo  {journal} {Sci. China
  Mater.}\ }\textbf {\bibinfo {volume} {62}},\ \bibinfo {pages} {1087}
  (\bibinfo {year} {2019})}\BibitemShut {NoStop}%
\bibitem [{\citenamefont {Chen}\ \emph {et~al.}(2017)\citenamefont {Chen},
  \citenamefont {Zhao}, \citenamefont {Tan}, \citenamefont {Xu}, \citenamefont
  {Wu}, \citenamefont {Liu}, \citenamefont {Fu}, \citenamefont {Fu},
  \citenamefont {Geng}, \citenamefont {Liu} \emph {et~al.}}]{Chen2017}%
  \BibitemOpen
  \bibfield  {author} {\bibinfo {author} {\bibfnamefont {J.}~\bibnamefont
  {Chen}}, \bibinfo {author} {\bibfnamefont {X.}~\bibnamefont {Zhao}}, \bibinfo
  {author} {\bibfnamefont {S.~J.}\ \bibnamefont {Tan}}, \bibinfo {author}
  {\bibfnamefont {H.}~\bibnamefont {Xu}}, \bibinfo {author} {\bibfnamefont
  {B.}~\bibnamefont {Wu}}, \bibinfo {author} {\bibfnamefont {B.}~\bibnamefont
  {Liu}}, \bibinfo {author} {\bibfnamefont {D.}~\bibnamefont {Fu}}, \bibinfo
  {author} {\bibfnamefont {W.}~\bibnamefont {Fu}}, \bibinfo {author}
  {\bibfnamefont {D.}~\bibnamefont {Geng}}, \bibinfo {author} {\bibfnamefont
  {Y.}~\bibnamefont {Liu}}, \emph {et~al.},\ }\bibfield  {title} {\bibinfo
  {title} {Chemical vapor deposition of large-size monolayer {MoSe$_2$}
  crystals on molten glass},\ }\href@noop {} {\bibfield  {journal} {\bibinfo
  {journal} {J. Am. Chem. Soc.}\ }\textbf {\bibinfo {volume} {139}},\ \bibinfo
  {pages} {1073} (\bibinfo {year} {2017})}\BibitemShut {NoStop}%
\bibitem [{\citenamefont {Zeng}\ \emph {et~al.}(2014)\citenamefont {Zeng},
  \citenamefont {Tan}, \citenamefont {Wang}, \citenamefont {Chen},
  \citenamefont {R{\"u}mmeli},\ and\ \citenamefont {Fu}}]{Zeng2014}%
  \BibitemOpen
  \bibfield  {author} {\bibinfo {author} {\bibfnamefont {M.}~\bibnamefont
  {Zeng}}, \bibinfo {author} {\bibfnamefont {L.}~\bibnamefont {Tan}}, \bibinfo
  {author} {\bibfnamefont {J.}~\bibnamefont {Wang}}, \bibinfo {author}
  {\bibfnamefont {L.}~\bibnamefont {Chen}}, \bibinfo {author} {\bibfnamefont
  {M.~H.}\ \bibnamefont {R{\"u}mmeli}},\ and\ \bibinfo {author} {\bibfnamefont
  {L.}~\bibnamefont {Fu}},\ }\bibfield  {title} {\bibinfo {title} {Liquid
  metal: an innovative solution to uniform graphene films},\ }\href@noop {}
  {\bibfield  {journal} {\bibinfo  {journal} {Chem. Mater.}\ }\textbf {\bibinfo
  {volume} {26}},\ \bibinfo {pages} {3637} (\bibinfo {year}
  {2014})}\BibitemShut {NoStop}%
\bibitem [{\citenamefont {Zeng}\ and\ \citenamefont {Fu}(2018)}]{Zeng2018}%
  \BibitemOpen
  \bibfield  {author} {\bibinfo {author} {\bibfnamefont {M.}~\bibnamefont
  {Zeng}}\ and\ \bibinfo {author} {\bibfnamefont {L.}~\bibnamefont {Fu}},\
  }\bibfield  {title} {\bibinfo {title} {Controllable fabrication of graphene
  and related two-dimensional materials on liquid metals via chemical vapor
  deposition},\ }\href {https://doi.org/10.1021/acs.accounts.8b00293}
  {\bibfield  {journal} {\bibinfo  {journal} {Acc. Mater. Res.}\ }\textbf
  {\bibinfo {volume} {51}},\ \bibinfo {pages} {2839} (\bibinfo {year}
  {2018})}\BibitemShut {NoStop}%
\bibitem [{\citenamefont {Xu}\ \emph {et~al.}(2020)\citenamefont {Xu},
  \citenamefont {Zhao}, \citenamefont {Qiu}, \citenamefont {Zhang},
  \citenamefont {Qiao},\ and\ \citenamefont {Ding}}]{Xu2020}%
  \BibitemOpen
  \bibfield  {author} {\bibinfo {author} {\bibfnamefont {Z.}~\bibnamefont
  {Xu}}, \bibinfo {author} {\bibfnamefont {G.}~\bibnamefont {Zhao}}, \bibinfo
  {author} {\bibfnamefont {L.}~\bibnamefont {Qiu}}, \bibinfo {author}
  {\bibfnamefont {X.}~\bibnamefont {Zhang}}, \bibinfo {author} {\bibfnamefont
  {G.}~\bibnamefont {Qiao}},\ and\ \bibinfo {author} {\bibfnamefont
  {F.}~\bibnamefont {Ding}},\ }\bibfield  {title} {\bibinfo {title} {Molecular
  dynamics simulation of graphene sinking during chemical vapor deposition
  growth on semi-molten cu substrate},\ }\href
  {https://doi.org/https://doi.org/10.1038/s41524-020-0281-1} {\bibfield
  {journal} {\bibinfo  {journal} {npj Comput. Mater.}\ }\textbf {\bibinfo
  {volume} {6}},\ \bibinfo {pages} {14} (\bibinfo {year} {2020})}\BibitemShut
  {NoStop}%
\bibitem [{\citenamefont {Schwarcz}\ and\ \citenamefont
  {Burov}(2022)}]{Burov2022}%
  \BibitemOpen
  \bibfield  {author} {\bibinfo {author} {\bibfnamefont {D.}~\bibnamefont
  {Schwarcz}}\ and\ \bibinfo {author} {\bibfnamefont {S.}~\bibnamefont
  {Burov}},\ }\bibfield  {title} {\bibinfo {title} {Self-assembly of
  two-dimensional, amorphous materials on a liquid substrate},\ }\href
  {https://doi.org/10.1103/PhysRevE.105.L022601} {\bibfield  {journal}
  {\bibinfo  {journal} {Phys. Rev. E}\ }\textbf {\bibinfo {volume} {105}},\
  \bibinfo {pages} {L022601} (\bibinfo {year} {2022})}\BibitemShut {NoStop}%
\bibitem [{\citenamefont {Elder}\ \emph {et~al.}(2002)\citenamefont {Elder},
  \citenamefont {Katakowski}, \citenamefont {Haataja},\ and\ \citenamefont
  {Grant}}]{Elder2002}%
  \BibitemOpen
  \bibfield  {author} {\bibinfo {author} {\bibfnamefont {K.~R.}\ \bibnamefont
  {Elder}}, \bibinfo {author} {\bibfnamefont {M.}~\bibnamefont {Katakowski}},
  \bibinfo {author} {\bibfnamefont {M.}~\bibnamefont {Haataja}},\ and\ \bibinfo
  {author} {\bibfnamefont {M.}~\bibnamefont {Grant}},\ }\bibfield  {title}
  {\bibinfo {title} {Modeling elasticity in crystal growth},\ }\href
  {https://doi.org/10.1103/PhysRevLett.88.245701} {\bibfield  {journal}
  {\bibinfo  {journal} {Phys. Rev. Lett.}\ }\textbf {\bibinfo {volume} {88}},\
  \bibinfo {pages} {245701} (\bibinfo {year} {2002})}\BibitemShut {NoStop}%
\bibitem [{\citenamefont {Elder}\ and\ \citenamefont
  {Grant}(2004)}]{Elder2004}%
  \BibitemOpen
  \bibfield  {author} {\bibinfo {author} {\bibfnamefont {K.~R.}\ \bibnamefont
  {Elder}}\ and\ \bibinfo {author} {\bibfnamefont {M.}~\bibnamefont {Grant}},\
  }\bibfield  {title} {\bibinfo {title} {Modeling elastic and plastic
  deformations in nonequilibrium processing using phase field crystals},\
  }\href {https://doi.org/10.1103/PhysRevE.70.051605} {\bibfield  {journal}
  {\bibinfo  {journal} {Phys. Rev. E}\ }\textbf {\bibinfo {volume} {70}},\
  \bibinfo {pages} {051605} (\bibinfo {year} {2004})}\BibitemShut {NoStop}%
\bibitem [{\citenamefont {Elder}\ \emph {et~al.}(2007)\citenamefont {Elder},
  \citenamefont {Provatas}, \citenamefont {Berry}, \citenamefont {Stefanovic},\
  and\ \citenamefont {Grant}}]{Elder2007}%
  \BibitemOpen
  \bibfield  {author} {\bibinfo {author} {\bibfnamefont {K.~R.}\ \bibnamefont
  {Elder}}, \bibinfo {author} {\bibfnamefont {N.}~\bibnamefont {Provatas}},
  \bibinfo {author} {\bibfnamefont {J.}~\bibnamefont {Berry}}, \bibinfo
  {author} {\bibfnamefont {P.}~\bibnamefont {Stefanovic}},\ and\ \bibinfo
  {author} {\bibfnamefont {M.}~\bibnamefont {Grant}},\ }\bibfield  {title}
  {\bibinfo {title} {Phase-field crystal modeling and classical density
  functional theory of freezing},\ }\href@noop {} {\bibfield  {journal}
  {\bibinfo  {journal} {Phys. Rev. B}\ }\textbf {\bibinfo {volume} {75}},\
  \bibinfo {pages} {064107} (\bibinfo {year} {2007})}\BibitemShut {NoStop}%
\bibitem [{\citenamefont {Greenwood}\ \emph {et~al.}(2010)\citenamefont
  {Greenwood}, \citenamefont {Provatas},\ and\ \citenamefont
  {Rottler}}]{Greenwood2010}%
  \BibitemOpen
  \bibfield  {author} {\bibinfo {author} {\bibfnamefont {M.}~\bibnamefont
  {Greenwood}}, \bibinfo {author} {\bibfnamefont {N.}~\bibnamefont
  {Provatas}},\ and\ \bibinfo {author} {\bibfnamefont {J.}~\bibnamefont
  {Rottler}},\ }\bibfield  {title} {\bibinfo {title} {Free energy functionals
  for efficient phase field crystal modeling of structural phase
  transformations},\ }\href@noop {} {\bibfield  {journal} {\bibinfo  {journal}
  {Phys. Rev. Lett.}\ }\textbf {\bibinfo {volume} {105}},\ \bibinfo {pages}
  {045702} (\bibinfo {year} {2010})}\BibitemShut {NoStop}%
\bibitem [{\citenamefont {Mkhonta}\ \emph {et~al.}(2013)\citenamefont
  {Mkhonta}, \citenamefont {Elder},\ and\ \citenamefont {Huang}}]{Mkhonta2013}%
  \BibitemOpen
  \bibfield  {author} {\bibinfo {author} {\bibfnamefont {S.~K.}\ \bibnamefont
  {Mkhonta}}, \bibinfo {author} {\bibfnamefont {K.~R.}\ \bibnamefont {Elder}},\
  and\ \bibinfo {author} {\bibfnamefont {Z.-F.}\ \bibnamefont {Huang}},\
  }\bibfield  {title} {\bibinfo {title} {Exploring the complex world of
  two-dimensional ordering with three modes},\ }\href@noop {} {\bibfield
  {journal} {\bibinfo  {journal} {Phys. Rev. Lett.}\ }\textbf {\bibinfo
  {volume} {111}},\ \bibinfo {pages} {035501} (\bibinfo {year}
  {2013})}\BibitemShut {NoStop}%
\bibitem [{\citenamefont {Huang}\ and\ \citenamefont
  {Elder}(2008)}]{Huang2008}%
  \BibitemOpen
  \bibfield  {author} {\bibinfo {author} {\bibfnamefont {Z.-F.}\ \bibnamefont
  {Huang}}\ and\ \bibinfo {author} {\bibfnamefont {K.~R.}\ \bibnamefont
  {Elder}},\ }\bibfield  {title} {\bibinfo {title} {Mesoscopic and microscopic
  modeling of island formation in strained film epitaxy},\ }\href@noop {}
  {\bibfield  {journal} {\bibinfo  {journal} {Phys. Rev. Lett.}\ }\textbf
  {\bibinfo {volume} {101}},\ \bibinfo {pages} {158701} (\bibinfo {year}
  {2008})}\BibitemShut {NoStop}%
\bibitem [{\citenamefont {Huang}\ and\ \citenamefont
  {Elder}(2010)}]{Huang2010}%
  \BibitemOpen
  \bibfield  {author} {\bibinfo {author} {\bibfnamefont {Z.-F.}\ \bibnamefont
  {Huang}}\ and\ \bibinfo {author} {\bibfnamefont {K.~R.}\ \bibnamefont
  {Elder}},\ }\bibfield  {title} {\bibinfo {title} {Morphological instability,
  evolution, and scaling in strained epitaxial films: An amplitude-equation
  analysis of the phase-field-crystal model},\ }\href@noop {} {\bibfield
  {journal} {\bibinfo  {journal} {Phys. Rev. B}\ }\textbf {\bibinfo {volume}
  {81}},\ \bibinfo {pages} {165421} (\bibinfo {year} {2010})}\BibitemShut
  {NoStop}%
\bibitem [{\citenamefont {Wu}\ and\ \citenamefont {Voorhees}(2009)}]{Wu2009}%
  \BibitemOpen
  \bibfield  {author} {\bibinfo {author} {\bibfnamefont {K.-A.}\ \bibnamefont
  {Wu}}\ and\ \bibinfo {author} {\bibfnamefont {P.~W.}\ \bibnamefont
  {Voorhees}},\ }\bibfield  {title} {\bibinfo {title} {Stress-induced
  morphological instabilities at the nanoscale examined using the phase field
  crystal approach},\ }\href {https://doi.org/10.1103/PhysRevB.80.125408}
  {\bibfield  {journal} {\bibinfo  {journal} {Phys. Rev. B}\ }\textbf {\bibinfo
  {volume} {80}},\ \bibinfo {pages} {125408} (\bibinfo {year}
  {2009})}\BibitemShut {NoStop}%
\bibitem [{\citenamefont {Elder}\ \emph {et~al.}(2012)\citenamefont {Elder},
  \citenamefont {Rossi}, \citenamefont {Kanerva}, \citenamefont {Sanches},
  \citenamefont {Ying}, \citenamefont {Granato}, \citenamefont {Achim},\ and\
  \citenamefont {Ala-Nissila}}]{Elder2012}%
  \BibitemOpen
  \bibfield  {author} {\bibinfo {author} {\bibfnamefont {K.~R.}\ \bibnamefont
  {Elder}}, \bibinfo {author} {\bibfnamefont {G.}~\bibnamefont {Rossi}},
  \bibinfo {author} {\bibfnamefont {P.}~\bibnamefont {Kanerva}}, \bibinfo
  {author} {\bibfnamefont {F.}~\bibnamefont {Sanches}}, \bibinfo {author}
  {\bibfnamefont {S.-C.}\ \bibnamefont {Ying}}, \bibinfo {author}
  {\bibfnamefont {E.}~\bibnamefont {Granato}}, \bibinfo {author} {\bibfnamefont
  {C.~V.}\ \bibnamefont {Achim}},\ and\ \bibinfo {author} {\bibfnamefont
  {T.}~\bibnamefont {Ala-Nissila}},\ }\bibfield  {title} {\bibinfo {title}
  {Patterning of heteroepitaxial overlayers from nano to micron scales},\
  }\href {https://doi.org/10.1103/PhysRevLett.108.226102} {\bibfield  {journal}
  {\bibinfo  {journal} {Phys. Rev. Lett.}\ }\textbf {\bibinfo {volume} {108}},\
  \bibinfo {pages} {226102} (\bibinfo {year} {2012})}\BibitemShut {NoStop}%
\bibitem [{\citenamefont {Elder}\ \emph {et~al.}(2016)\citenamefont {Elder},
  \citenamefont {Chen}, \citenamefont {Elder}, \citenamefont {Hirvonen},
  \citenamefont {Mkhonta}, \citenamefont {Ying}, \citenamefont {Granato},
  \citenamefont {Huang},\ and\ \citenamefont {Ala-Nissila}}]{Elder2016}%
  \BibitemOpen
  \bibfield  {author} {\bibinfo {author} {\bibfnamefont {K.~R.}\ \bibnamefont
  {Elder}}, \bibinfo {author} {\bibfnamefont {Z.}~\bibnamefont {Chen}},
  \bibinfo {author} {\bibfnamefont {K.~L.~M.}\ \bibnamefont {Elder}}, \bibinfo
  {author} {\bibfnamefont {P.}~\bibnamefont {Hirvonen}}, \bibinfo {author}
  {\bibfnamefont {S.~K.}\ \bibnamefont {Mkhonta}}, \bibinfo {author}
  {\bibfnamefont {S.-C.}\ \bibnamefont {Ying}}, \bibinfo {author}
  {\bibfnamefont {E.}~\bibnamefont {Granato}}, \bibinfo {author} {\bibfnamefont
  {Z.-F.}\ \bibnamefont {Huang}},\ and\ \bibinfo {author} {\bibfnamefont
  {T.}~\bibnamefont {Ala-Nissila}},\ }\bibfield  {title} {\bibinfo {title}
  {{Honeycomb and triangular domain wall networks in heteroepitaxial
  systems}},\ }\href {https://doi.org/10.1063/1.4948370} {\bibfield  {journal}
  {\bibinfo  {journal} {J. Chem. Phys.}\ }\textbf {\bibinfo {volume} {144}},\
  \bibinfo {pages} {174703} (\bibinfo {year} {2016})}\BibitemShut {NoStop}%
\bibitem [{\citenamefont {Granato}\ \emph {et~al.}(2011)\citenamefont
  {Granato}, \citenamefont {Ramos}, \citenamefont {Achim}, \citenamefont
  {Lehikoinen}, \citenamefont {Ying}, \citenamefont {Ala-Nissila},\ and\
  \citenamefont {Elder}}]{Granato2011}%
  \BibitemOpen
  \bibfield  {author} {\bibinfo {author} {\bibfnamefont {E.}~\bibnamefont
  {Granato}}, \bibinfo {author} {\bibfnamefont {J.~A.~P.}\ \bibnamefont
  {Ramos}}, \bibinfo {author} {\bibfnamefont {C.~V.}\ \bibnamefont {Achim}},
  \bibinfo {author} {\bibfnamefont {J.}~\bibnamefont {Lehikoinen}}, \bibinfo
  {author} {\bibfnamefont {S.~C.}\ \bibnamefont {Ying}}, \bibinfo {author}
  {\bibfnamefont {T.}~\bibnamefont {Ala-Nissila}},\ and\ \bibinfo {author}
  {\bibfnamefont {K.~R.}\ \bibnamefont {Elder}},\ }\bibfield  {title} {\bibinfo
  {title} {Glassy phases and driven response of the phase-field-crystal model
  with random pinning},\ }\href {https://doi.org/10.1103/PhysRevE.84.031102}
  {\bibfield  {journal} {\bibinfo  {journal} {Phys. Rev. E}\ }\textbf {\bibinfo
  {volume} {84}},\ \bibinfo {pages} {031102} (\bibinfo {year}
  {2011})}\BibitemShut {NoStop}%
\bibitem [{\citenamefont {Torquato}\ and\ \citenamefont
  {Stillinger}(2003)}]{Torquato2003}%
  \BibitemOpen
  \bibfield  {author} {\bibinfo {author} {\bibfnamefont {S.}~\bibnamefont
  {Torquato}}\ and\ \bibinfo {author} {\bibfnamefont {F.~H.}\ \bibnamefont
  {Stillinger}},\ }\bibfield  {title} {\bibinfo {title} {Local density
  fluctuations, hyperuniformity, and order metrics},\ }\href
  {https://doi.org/10.1103/PhysRevE.68.041113} {\bibfield  {journal} {\bibinfo
  {journal} {Phys. Rev. E}\ }\textbf {\bibinfo {volume} {68}},\ \bibinfo
  {pages} {041113} (\bibinfo {year} {2003})}\BibitemShut {NoStop}%
\bibitem [{\citenamefont {Wilken}\ \emph {et~al.}(2023)\citenamefont {Wilken},
  \citenamefont {Chaderjian},\ and\ \citenamefont {Saleh}}]{Wilken2023}%
  \BibitemOpen
  \bibfield  {author} {\bibinfo {author} {\bibfnamefont {S.}~\bibnamefont
  {Wilken}}, \bibinfo {author} {\bibfnamefont {A.}~\bibnamefont {Chaderjian}},\
  and\ \bibinfo {author} {\bibfnamefont {O.~A.}\ \bibnamefont {Saleh}},\
  }\bibfield  {title} {\bibinfo {title} {Spatial organization of
  phase-separated {DNA} droplets},\ }\href
  {https://doi.org/10.1103/PhysRevX.13.031014} {\bibfield  {journal} {\bibinfo
  {journal} {Phys. Rev. X}\ }\textbf {\bibinfo {volume} {13}},\ \bibinfo
  {pages} {031014} (\bibinfo {year} {2023})}\BibitemShut {NoStop}%
\bibitem [{\citenamefont {Zito}\ \emph {et~al.}(2015)\citenamefont {Zito},
  \citenamefont {Rusciano}, \citenamefont {Pesce}, \citenamefont {Malafronte},
  \citenamefont {Di~Girolamo}, \citenamefont {Ausanio}, \citenamefont
  {Vecchione},\ and\ \citenamefont {Sasso}}]{Zito2015}%
  \BibitemOpen
  \bibfield  {author} {\bibinfo {author} {\bibfnamefont {G.}~\bibnamefont
  {Zito}}, \bibinfo {author} {\bibfnamefont {G.}~\bibnamefont {Rusciano}},
  \bibinfo {author} {\bibfnamefont {G.}~\bibnamefont {Pesce}}, \bibinfo
  {author} {\bibfnamefont {A.}~\bibnamefont {Malafronte}}, \bibinfo {author}
  {\bibfnamefont {R.}~\bibnamefont {Di~Girolamo}}, \bibinfo {author}
  {\bibfnamefont {G.}~\bibnamefont {Ausanio}}, \bibinfo {author} {\bibfnamefont
  {A.}~\bibnamefont {Vecchione}},\ and\ \bibinfo {author} {\bibfnamefont
  {A.}~\bibnamefont {Sasso}},\ }\bibfield  {title} {\bibinfo {title} {Nanoscale
  engineering of two-dimensional disordered hyperuniform block-copolymer
  assemblies},\ }\href {https://doi.org/10.1103/PhysRevE.92.050601} {\bibfield
  {journal} {\bibinfo  {journal} {Phys. Rev. E}\ }\textbf {\bibinfo {volume}
  {92}},\ \bibinfo {pages} {050601} (\bibinfo {year} {2015})}\BibitemShut
  {NoStop}%
\bibitem [{\citenamefont {Chremos}\ and\ \citenamefont
  {Douglas}(2018)}]{Chremos2018}%
  \BibitemOpen
  \bibfield  {author} {\bibinfo {author} {\bibfnamefont {A.}~\bibnamefont
  {Chremos}}\ and\ \bibinfo {author} {\bibfnamefont {J.~F.}\ \bibnamefont
  {Douglas}},\ }\bibfield  {title} {\bibinfo {title} {Hidden hyperuniformity in
  soft polymeric materials},\ }\href
  {https://doi.org/10.1103/PhysRevLett.121.258002} {\bibfield  {journal}
  {\bibinfo  {journal} {Phys. Rev. Lett.}\ }\textbf {\bibinfo {volume} {121}},\
  \bibinfo {pages} {258002} (\bibinfo {year} {2018})}\BibitemShut {NoStop}%
\bibitem [{\citenamefont {Chremos}(2020)}]{Chremos2020}%
  \BibitemOpen
  \bibfield  {author} {\bibinfo {author} {\bibfnamefont {A.}~\bibnamefont
  {Chremos}},\ }\bibfield  {title} {\bibinfo {title} {{Design of nearly perfect
  hyperuniform polymeric materials}},\ }\href
  {https://doi.org/10.1063/5.0017861} {\bibfield  {journal} {\bibinfo
  {journal} {J. Chem. Phys.}\ }\textbf {\bibinfo {volume} {153}},\ \bibinfo
  {pages} {054902} (\bibinfo {year} {2020})}\BibitemShut {NoStop}%
\bibitem [{\citenamefont {Yu}\ \emph {et~al.}(2021)\citenamefont {Yu},
  \citenamefont {Qiu}, \citenamefont {Chong}, \citenamefont {Torquato},\ and\
  \citenamefont {Park}}]{Yu2021}%
  \BibitemOpen
  \bibfield  {author} {\bibinfo {author} {\bibfnamefont {S.}~\bibnamefont
  {Yu}}, \bibinfo {author} {\bibfnamefont {C.-W.}\ \bibnamefont {Qiu}},
  \bibinfo {author} {\bibfnamefont {Y.}~\bibnamefont {Chong}}, \bibinfo
  {author} {\bibfnamefont {S.}~\bibnamefont {Torquato}},\ and\ \bibinfo
  {author} {\bibfnamefont {N.}~\bibnamefont {Park}},\ }\bibfield  {title}
  {\bibinfo {title} {Engineered disorder in photonics},\ }\href@noop {}
  {\bibfield  {journal} {\bibinfo  {journal} {Nat. Rev. Mater.}\ }\textbf
  {\bibinfo {volume} {6}},\ \bibinfo {pages} {226} (\bibinfo {year}
  {2021})}\BibitemShut {NoStop}%
\bibitem [{\citenamefont {Man}\ \emph {et~al.}(2013)\citenamefont {Man},
  \citenamefont {Florescu}, \citenamefont {Williamson}, \citenamefont {He},
  \citenamefont {Hashemizad}, \citenamefont {Leung}, \citenamefont {Liner},
  \citenamefont {Torquato}, \citenamefont {Chaikin},\ and\ \citenamefont
  {Steinhardt}}]{Man2013}%
  \BibitemOpen
  \bibfield  {author} {\bibinfo {author} {\bibfnamefont {W.}~\bibnamefont
  {Man}}, \bibinfo {author} {\bibfnamefont {M.}~\bibnamefont {Florescu}},
  \bibinfo {author} {\bibfnamefont {E.~P.}\ \bibnamefont {Williamson}},
  \bibinfo {author} {\bibfnamefont {Y.}~\bibnamefont {He}}, \bibinfo {author}
  {\bibfnamefont {S.~R.}\ \bibnamefont {Hashemizad}}, \bibinfo {author}
  {\bibfnamefont {B.~Y.}\ \bibnamefont {Leung}}, \bibinfo {author}
  {\bibfnamefont {D.~R.}\ \bibnamefont {Liner}}, \bibinfo {author}
  {\bibfnamefont {S.}~\bibnamefont {Torquato}}, \bibinfo {author}
  {\bibfnamefont {P.~M.}\ \bibnamefont {Chaikin}},\ and\ \bibinfo {author}
  {\bibfnamefont {P.~J.}\ \bibnamefont {Steinhardt}},\ }\bibfield  {title}
  {\bibinfo {title} {Isotropic band gaps and freeform waveguides observed in
  hyperuniform disordered photonic solids},\ }\href@noop {} {\bibfield
  {journal} {\bibinfo  {journal} {Proc. Natl. Acad. Sci. U.S.A.}\ }\textbf
  {\bibinfo {volume} {110}},\ \bibinfo {pages} {15886} (\bibinfo {year}
  {2013})}\BibitemShut {NoStop}%
\bibitem [{\citenamefont {Oskooi}\ \emph {et~al.}(2012)\citenamefont {Oskooi},
  \citenamefont {Favuzzi}, \citenamefont {Tanaka}, \citenamefont {Shigeta},
  \citenamefont {Kawakami},\ and\ \citenamefont {Noda}}]{Oskooi2012}%
  \BibitemOpen
  \bibfield  {author} {\bibinfo {author} {\bibfnamefont {A.}~\bibnamefont
  {Oskooi}}, \bibinfo {author} {\bibfnamefont {P.~A.}\ \bibnamefont {Favuzzi}},
  \bibinfo {author} {\bibfnamefont {Y.}~\bibnamefont {Tanaka}}, \bibinfo
  {author} {\bibfnamefont {H.}~\bibnamefont {Shigeta}}, \bibinfo {author}
  {\bibfnamefont {Y.}~\bibnamefont {Kawakami}},\ and\ \bibinfo {author}
  {\bibfnamefont {S.}~\bibnamefont {Noda}},\ }\bibfield  {title} {\bibinfo
  {title} {Partially disordered photonic-crystal thin films for enhanced and
  robust photovoltaics},\ }\href@noop {} {\bibfield  {journal} {\bibinfo
  {journal} {Appl. Phys. Lett.}\ }\textbf {\bibinfo {volume} {100}} (\bibinfo
  {year} {2012})}\BibitemShut {NoStop}%
\bibitem [{\citenamefont {Pala}\ \emph {et~al.}(2013)\citenamefont {Pala},
  \citenamefont {Liu}, \citenamefont {Barnard}, \citenamefont {Askarov},
  \citenamefont {Garnett}, \citenamefont {Fan},\ and\ \citenamefont
  {Brongersma}}]{Pala2013}%
  \BibitemOpen
  \bibfield  {author} {\bibinfo {author} {\bibfnamefont {R.~A.}\ \bibnamefont
  {Pala}}, \bibinfo {author} {\bibfnamefont {J.~S.}\ \bibnamefont {Liu}},
  \bibinfo {author} {\bibfnamefont {E.~S.}\ \bibnamefont {Barnard}}, \bibinfo
  {author} {\bibfnamefont {D.}~\bibnamefont {Askarov}}, \bibinfo {author}
  {\bibfnamefont {E.~C.}\ \bibnamefont {Garnett}}, \bibinfo {author}
  {\bibfnamefont {S.}~\bibnamefont {Fan}},\ and\ \bibinfo {author}
  {\bibfnamefont {M.~L.}\ \bibnamefont {Brongersma}},\ }\bibfield  {title}
  {\bibinfo {title} {Optimization of non-periodic plasmonic light-trapping
  layers for thin-film solar cells},\ }\href@noop {} {\bibfield  {journal}
  {\bibinfo  {journal} {Nat. Commun.}\ }\textbf {\bibinfo {volume} {4}},\
  \bibinfo {pages} {2095} (\bibinfo {year} {2013})}\BibitemShut {NoStop}%
\bibitem [{\citenamefont {Le~Thien}\ \emph {et~al.}(2017)\citenamefont
  {Le~Thien}, \citenamefont {McDermott}, \citenamefont {Reichhardt},\ and\
  \citenamefont {Reichhardt}}]{Thien2017}%
  \BibitemOpen
  \bibfield  {author} {\bibinfo {author} {\bibfnamefont {Q.}~\bibnamefont
  {Le~Thien}}, \bibinfo {author} {\bibfnamefont {D.}~\bibnamefont {McDermott}},
  \bibinfo {author} {\bibfnamefont {C.~J.~O.}\ \bibnamefont {Reichhardt}},\
  and\ \bibinfo {author} {\bibfnamefont {C.}~\bibnamefont {Reichhardt}},\
  }\bibfield  {title} {\bibinfo {title} {Enhanced pinning for vortices in
  hyperuniform pinning arrays and emergent hyperuniform vortex configurations
  with quenched disorder},\ }\href {https://doi.org/10.1103/PhysRevB.96.094516}
  {\bibfield  {journal} {\bibinfo  {journal} {Phys. Rev. B}\ }\textbf {\bibinfo
  {volume} {96}},\ \bibinfo {pages} {094516} (\bibinfo {year}
  {2017})}\BibitemShut {NoStop}%
\bibitem [{\citenamefont {Llorens}\ \emph {et~al.}(2020)\citenamefont
  {Llorens}, \citenamefont {Guillam\'on}, \citenamefont {Serrano},
  \citenamefont {C\'ordoba}, \citenamefont {Ses\'e}, \citenamefont {De~Teresa},
  \citenamefont {Ibarra}, \citenamefont {Vieira}, \citenamefont {Ortu\~no},\
  and\ \citenamefont {Suderow}}]{Llorens2020}%
  \BibitemOpen
  \bibfield  {author} {\bibinfo {author} {\bibfnamefont {J.~B.}\ \bibnamefont
  {Llorens}}, \bibinfo {author} {\bibfnamefont {I.}~\bibnamefont
  {Guillam\'on}}, \bibinfo {author} {\bibfnamefont {I.~G.}\ \bibnamefont
  {Serrano}}, \bibinfo {author} {\bibfnamefont {R.}~\bibnamefont {C\'ordoba}},
  \bibinfo {author} {\bibfnamefont {J.}~\bibnamefont {Ses\'e}}, \bibinfo
  {author} {\bibfnamefont {J.~M.}\ \bibnamefont {De~Teresa}}, \bibinfo {author}
  {\bibfnamefont {M.~R.}\ \bibnamefont {Ibarra}}, \bibinfo {author}
  {\bibfnamefont {S.}~\bibnamefont {Vieira}}, \bibinfo {author} {\bibfnamefont
  {M.}~\bibnamefont {Ortu\~no}},\ and\ \bibinfo {author} {\bibfnamefont
  {H.}~\bibnamefont {Suderow}},\ }\bibfield  {title} {\bibinfo {title}
  {Disordered hyperuniformity in superconducting vortex lattices},\ }\href
  {https://doi.org/10.1103/PhysRevResearch.2.033133} {\bibfield  {journal}
  {\bibinfo  {journal} {Phys. Rev. Res.}\ }\textbf {\bibinfo {volume} {2}},\
  \bibinfo {pages} {033133} (\bibinfo {year} {2020})}\BibitemShut {NoStop}%
\bibitem [{\citenamefont {Radzihovsky}\ and\ \citenamefont
  {Toner}(1997)}]{Toner1997}%
  \BibitemOpen
  \bibfield  {author} {\bibinfo {author} {\bibfnamefont {L.}~\bibnamefont
  {Radzihovsky}}\ and\ \bibinfo {author} {\bibfnamefont {J.}~\bibnamefont
  {Toner}},\ }\bibfield  {title} {\bibinfo {title}
  {Nematic--to--smectic-$\mathit{A}$ transition in aerogel},\ }\href
  {https://doi.org/10.1103/PhysRevLett.79.4214} {\bibfield  {journal} {\bibinfo
   {journal} {Phys. Rev. Lett.}\ }\textbf {\bibinfo {volume} {79}},\ \bibinfo
  {pages} {4214} (\bibinfo {year} {1997})}\BibitemShut {NoStop}%
\bibitem [{\citenamefont {Radzihovsky}\ and\ \citenamefont
  {Toner}(1999)}]{Toner1999}%
  \BibitemOpen
  \bibfield  {author} {\bibinfo {author} {\bibfnamefont {L.}~\bibnamefont
  {Radzihovsky}}\ and\ \bibinfo {author} {\bibfnamefont {J.}~\bibnamefont
  {Toner}},\ }\bibfield  {title} {\bibinfo {title} {Smectic liquid crystals in
  random environments},\ }\href {https://doi.org/10.1103/PhysRevB.60.206}
  {\bibfield  {journal} {\bibinfo  {journal} {Phys. Rev. B}\ }\textbf {\bibinfo
  {volume} {60}},\ \bibinfo {pages} {206} (\bibinfo {year} {1999})}\BibitemShut
  {NoStop}%
\bibitem [{\citenamefont {Bellini}\ \emph {et~al.}(2001)\citenamefont
  {Bellini}, \citenamefont {Radzihovsky}, \citenamefont {Toner},\ and\
  \citenamefont {Clark}}]{Bellini2001}%
  \BibitemOpen
  \bibfield  {author} {\bibinfo {author} {\bibfnamefont {T.}~\bibnamefont
  {Bellini}}, \bibinfo {author} {\bibfnamefont {L.}~\bibnamefont
  {Radzihovsky}}, \bibinfo {author} {\bibfnamefont {J.}~\bibnamefont {Toner}},\
  and\ \bibinfo {author} {\bibfnamefont {N.~A.}\ \bibnamefont {Clark}},\
  }\bibfield  {title} {\bibinfo {title} {Universality and scaling in the
  disordering of a smectic liquid crystal},\ }\href
  {https://doi.org/10.1126/science.1057480} {\bibfield  {journal} {\bibinfo
  {journal} {Science}\ }\textbf {\bibinfo {volume} {294}},\ \bibinfo {pages}
  {1074} (\bibinfo {year} {2001})}\BibitemShut {NoStop}%
\bibitem [{\citenamefont {Ma}\ \emph {et~al.}(2023)\citenamefont {Ma},
  \citenamefont {Pausch},\ and\ \citenamefont {Cates}}]{Ma2023}%
  \BibitemOpen
  \bibfield  {author} {\bibinfo {author} {\bibfnamefont {X.}~\bibnamefont
  {Ma}}, \bibinfo {author} {\bibfnamefont {J.}~\bibnamefont {Pausch}},\ and\
  \bibinfo {author} {\bibfnamefont {M.~E.}\ \bibnamefont {Cates}},\ }\bibfield
  {title} {\bibinfo {title} {Theory of hyperuniformity at the absorbing state
  transition},\ }\href@noop {} {\bibfield  {journal} {\bibinfo  {journal}
  {arXiv preprint arXiv:2310.17391}\ } (\bibinfo {year} {2023})}\BibitemShut
  {NoStop}%
\bibitem [{\citenamefont {Hexner}\ and\ \citenamefont
  {Levine}(2017)}]{Hexner2017}%
  \BibitemOpen
  \bibfield  {author} {\bibinfo {author} {\bibfnamefont {D.}~\bibnamefont
  {Hexner}}\ and\ \bibinfo {author} {\bibfnamefont {D.}~\bibnamefont
  {Levine}},\ }\bibfield  {title} {\bibinfo {title} {Noise, diffusion, and
  hyperuniformity},\ }\href {https://doi.org/10.1103/PhysRevLett.118.020601}
  {\bibfield  {journal} {\bibinfo  {journal} {Phys. Rev. Lett.}\ }\textbf
  {\bibinfo {volume} {118}},\ \bibinfo {pages} {020601} (\bibinfo {year}
  {2017})}\BibitemShut {NoStop}%
\bibitem [{\citenamefont {Cheng}\ and\ \citenamefont
  {Warren}(2008)}]{Warren2008}%
  \BibitemOpen
  \bibfield  {author} {\bibinfo {author} {\bibfnamefont {M.}~\bibnamefont
  {Cheng}}\ and\ \bibinfo {author} {\bibfnamefont {J.~A.}\ \bibnamefont
  {Warren}},\ }\bibfield  {title} {\bibinfo {title} {An efficient algorithm for
  solving the phase field crystal model},\ }\href
  {https://doi.org/https://doi.org/10.1016/j.jcp.2008.03.012} {\bibfield
  {journal} {\bibinfo  {journal} {J. Comput. Phys.}\ }\textbf {\bibinfo
  {volume} {227}},\ \bibinfo {pages} {6241} (\bibinfo {year}
  {2008})}\BibitemShut {NoStop}%
\bibitem [{\citenamefont {Chen}\ \emph {et~al.}(2021)\citenamefont {Chen},
  \citenamefont {Zheng}, \citenamefont {Liu}, \citenamefont {Zhang},
  \citenamefont {Chen}, \citenamefont {Jiao},\ and\ \citenamefont
  {Zhuang}}]{Chen2020}%
  \BibitemOpen
  \bibfield  {author} {\bibinfo {author} {\bibfnamefont {D.}~\bibnamefont
  {Chen}}, \bibinfo {author} {\bibfnamefont {Y.}~\bibnamefont {Zheng}},
  \bibinfo {author} {\bibfnamefont {L.}~\bibnamefont {Liu}}, \bibinfo {author}
  {\bibfnamefont {G.}~\bibnamefont {Zhang}}, \bibinfo {author} {\bibfnamefont
  {M.}~\bibnamefont {Chen}}, \bibinfo {author} {\bibfnamefont {Y.}~\bibnamefont
  {Jiao}},\ and\ \bibinfo {author} {\bibfnamefont {H.}~\bibnamefont {Zhuang}},\
  }\bibfield  {title} {\bibinfo {title} {Stone-wales defects preserve
  hyperuniformity in amorphous two-dimensional networks},\ }\href@noop {}
  {\bibfield  {journal} {\bibinfo  {journal} {Proc. Natl. Acad. Sci. U.S.A.}\
  }\textbf {\bibinfo {volume} {118}},\ \bibinfo {pages} {e2016862118} (\bibinfo
  {year} {2021})}\BibitemShut {NoStop}%
\bibitem [{\citenamefont {Atkinson}\ \emph {et~al.}(2016)\citenamefont
  {Atkinson}, \citenamefont {Zhang}, \citenamefont {Hopkins},\ and\
  \citenamefont {Torquato}}]{Torquato2016}%
  \BibitemOpen
  \bibfield  {author} {\bibinfo {author} {\bibfnamefont {S.}~\bibnamefont
  {Atkinson}}, \bibinfo {author} {\bibfnamefont {G.}~\bibnamefont {Zhang}},
  \bibinfo {author} {\bibfnamefont {A.~B.}\ \bibnamefont {Hopkins}},\ and\
  \bibinfo {author} {\bibfnamefont {S.}~\bibnamefont {Torquato}},\ }\bibfield
  {title} {\bibinfo {title} {Critical slowing down and hyperuniformity on
  approach to jamming},\ }\href {https://doi.org/10.1103/PhysRevE.94.012902}
  {\bibfield  {journal} {\bibinfo  {journal} {Phys. Rev. E}\ }\textbf {\bibinfo
  {volume} {94}},\ \bibinfo {pages} {012902} (\bibinfo {year}
  {2016})}\BibitemShut {NoStop}%
\bibitem [{\citenamefont {Cross}\ and\ \citenamefont
  {Meiron}(1995)}]{Cross1995}%
  \BibitemOpen
  \bibfield  {author} {\bibinfo {author} {\bibfnamefont {M.~C.}\ \bibnamefont
  {Cross}}\ and\ \bibinfo {author} {\bibfnamefont {D.~I.}\ \bibnamefont
  {Meiron}},\ }\bibfield  {title} {\bibinfo {title} {Domain coarsening in
  systems far from equilibrium},\ }\href
  {https://doi.org/10.1103/PhysRevLett.75.2152} {\bibfield  {journal} {\bibinfo
   {journal} {Phys. Rev. Lett.}\ }\textbf {\bibinfo {volume} {75}},\ \bibinfo
  {pages} {2152} (\bibinfo {year} {1995})}\BibitemShut {NoStop}%
\bibitem [{\citenamefont {Hou}\ \emph {et~al.}(1997)\citenamefont {Hou},
  \citenamefont {Sasa},\ and\ \citenamefont {Goldenfeld}}]{Hou1997}%
  \BibitemOpen
  \bibfield  {author} {\bibinfo {author} {\bibfnamefont {Q.}~\bibnamefont
  {Hou}}, \bibinfo {author} {\bibfnamefont {S.}~\bibnamefont {Sasa}},\ and\
  \bibinfo {author} {\bibfnamefont {N.}~\bibnamefont {Goldenfeld}},\ }\bibfield
   {title} {\bibinfo {title} {Dynamical scaling behavior of the
  {Swift-Hohenberg} equation following a quench to the modulated state},\
  }\href {https://doi.org/https://doi.org/10.1016/S0378-4371(96)00480-3}
  {\bibfield  {journal} {\bibinfo  {journal} {Physica A}\ }\textbf {\bibinfo
  {volume} {239}},\ \bibinfo {pages} {219} (\bibinfo {year}
  {1997})}\BibitemShut {NoStop}%
\bibitem [{\citenamefont {Boyer}\ and\ \citenamefont
  {Vi\~nals}(2002)}]{Boyer2002}%
  \BibitemOpen
  \bibfield  {author} {\bibinfo {author} {\bibfnamefont {D.}~\bibnamefont
  {Boyer}}\ and\ \bibinfo {author} {\bibfnamefont {J.}~\bibnamefont
  {Vi\~nals}},\ }\bibfield  {title} {\bibinfo {title} {Grain boundary pinning
  and glassy dynamics in stripe phases},\ }\href
  {https://doi.org/10.1103/PhysRevE.65.046119} {\bibfield  {journal} {\bibinfo
  {journal} {Phys. Rev. E}\ }\textbf {\bibinfo {volume} {65}},\ \bibinfo
  {pages} {046119} (\bibinfo {year} {2002})}\BibitemShut {NoStop}%
\bibitem [{\citenamefont {Ohnogi}\ and\ \citenamefont
  {Shiwa}(2011)}]{Shiwa2011}%
  \BibitemOpen
  \bibfield  {author} {\bibinfo {author} {\bibfnamefont {H.}~\bibnamefont
  {Ohnogi}}\ and\ \bibinfo {author} {\bibfnamefont {Y.}~\bibnamefont {Shiwa}},\
  }\bibfield  {title} {\bibinfo {title} {Effect of noise on ordering of
  hexagonal grains in a phase-field-crystal model},\ }\href
  {https://doi.org/10.1103/PhysRevE.84.051603} {\bibfield  {journal} {\bibinfo
  {journal} {Phys. Rev. E}\ }\textbf {\bibinfo {volume} {84}},\ \bibinfo
  {pages} {051603} (\bibinfo {year} {2011})}\BibitemShut {NoStop}%
\bibitem [{\citenamefont {Hutchinson}\ \emph {et~al.}(2022)\citenamefont
  {Hutchinson}, \citenamefont {Lavergne},\ and\ \citenamefont
  {Dullens}}]{Hutchinson2022}%
  \BibitemOpen
  \bibfield  {author} {\bibinfo {author} {\bibfnamefont {J.~D.}\ \bibnamefont
  {Hutchinson}}, \bibinfo {author} {\bibfnamefont {F.~A.}\ \bibnamefont
  {Lavergne}},\ and\ \bibinfo {author} {\bibfnamefont {R.~P.~A.}\ \bibnamefont
  {Dullens}},\ }\bibfield  {title} {\bibinfo {title} {Crystallization and grain
  growth in impurity-doped colloidal polycrystals},\ }\href
  {https://doi.org/10.1103/PhysRevMaterials.6.075604} {\bibfield  {journal}
  {\bibinfo  {journal} {Phys. Rev. Mater.}\ }\textbf {\bibinfo {volume} {6}},\
  \bibinfo {pages} {075604} (\bibinfo {year} {2022})}\BibitemShut {NoStop}%
\bibitem [{\citenamefont {Tegze}\ \emph {et~al.}(2011)\citenamefont {Tegze},
  \citenamefont {T\'oth},\ and\ \citenamefont {Gr\'an\'asy}}]{Laszlo2011}%
  \BibitemOpen
  \bibfield  {author} {\bibinfo {author} {\bibfnamefont {G.}~\bibnamefont
  {Tegze}}, \bibinfo {author} {\bibfnamefont {G.~I.}\ \bibnamefont {T\'oth}},\
  and\ \bibinfo {author} {\bibfnamefont {L.}~\bibnamefont {Gr\'an\'asy}},\
  }\bibfield  {title} {\bibinfo {title} {Faceting and branching in {2D} crystal
  growth},\ }\href {https://doi.org/10.1103/PhysRevLett.106.195502} {\bibfield
  {journal} {\bibinfo  {journal} {Phys. Rev. Lett.}\ }\textbf {\bibinfo
  {volume} {106}},\ \bibinfo {pages} {195502} (\bibinfo {year}
  {2011})}\BibitemShut {NoStop}%
\bibitem [{\citenamefont {Granasy}\ \emph {et~al.}(2011)\citenamefont
  {Granasy}, \citenamefont {Tegze}, \citenamefont {Toth},\ and\ \citenamefont
  {Pusztai}}]{Laszlo2011b}%
  \BibitemOpen
  \bibfield  {author} {\bibinfo {author} {\bibfnamefont {L.}~\bibnamefont
  {Granasy}}, \bibinfo {author} {\bibfnamefont {G.}~\bibnamefont {Tegze}},
  \bibinfo {author} {\bibfnamefont {G.~I.}\ \bibnamefont {Toth}},\ and\
  \bibinfo {author} {\bibfnamefont {T.}~\bibnamefont {Pusztai}},\ }\bibfield
  {title} {\bibinfo {title} {Phase-field crystal modelling of crystal
  nucleation, heteroepitaxy and patterning},\ }\href@noop {} {\bibfield
  {journal} {\bibinfo  {journal} {Philos. Mag.}\ }\textbf {\bibinfo {volume}
  {91}},\ \bibinfo {pages} {123} (\bibinfo {year} {2011})}\BibitemShut
  {NoStop}%
\bibitem [{\citenamefont {Vargo}\ \emph {et~al.}(2023)\citenamefont {Vargo},
  \citenamefont {Ma}, \citenamefont {Li}, \citenamefont {Zhang}, \citenamefont
  {Kwon}, \citenamefont {Evans}, \citenamefont {Tang}, \citenamefont
  {Tovmasyan}, \citenamefont {Jan}, \citenamefont {Arias} \emph
  {et~al.}}]{Vargo2023}%
  \BibitemOpen
  \bibfield  {author} {\bibinfo {author} {\bibfnamefont {E.}~\bibnamefont
  {Vargo}}, \bibinfo {author} {\bibfnamefont {L.}~\bibnamefont {Ma}}, \bibinfo
  {author} {\bibfnamefont {H.}~\bibnamefont {Li}}, \bibinfo {author}
  {\bibfnamefont {Q.}~\bibnamefont {Zhang}}, \bibinfo {author} {\bibfnamefont
  {J.}~\bibnamefont {Kwon}}, \bibinfo {author} {\bibfnamefont {K.~M.}\
  \bibnamefont {Evans}}, \bibinfo {author} {\bibfnamefont {X.}~\bibnamefont
  {Tang}}, \bibinfo {author} {\bibfnamefont {V.~L.}\ \bibnamefont {Tovmasyan}},
  \bibinfo {author} {\bibfnamefont {J.}~\bibnamefont {Jan}}, \bibinfo {author}
  {\bibfnamefont {A.~C.}\ \bibnamefont {Arias}}, \emph {et~al.},\ }\bibfield
  {title} {\bibinfo {title} {Functional composites by programming
  entropy-driven nanosheet growth},\ }\href
  {https://doi.org/10.1038/s41586-023-06660-x} {\bibfield  {journal} {\bibinfo
  {journal} {Nature}\ }\textbf {\bibinfo {volume} {623}},\ \bibinfo {pages}
  {724} (\bibinfo {year} {2023})}\BibitemShut {NoStop}%
\bibitem [{\citenamefont {Bruevich}\ \emph {et~al.}(2019)\citenamefont
  {Bruevich}, \citenamefont {Glushkova}, \citenamefont {Poimanova},
  \citenamefont {Fedorenko}, \citenamefont {Luponosov}, \citenamefont
  {Bakirov}, \citenamefont {Shcherbina}, \citenamefont {Chvalun}, \citenamefont
  {Sosorev}, \citenamefont {Grodd} \emph {et~al.}}]{Bruevich2019}%
  \BibitemOpen
  \bibfield  {author} {\bibinfo {author} {\bibfnamefont {V.~V.}\ \bibnamefont
  {Bruevich}}, \bibinfo {author} {\bibfnamefont {A.~V.}\ \bibnamefont
  {Glushkova}}, \bibinfo {author} {\bibfnamefont {O.~Y.}\ \bibnamefont
  {Poimanova}}, \bibinfo {author} {\bibfnamefont {R.~S.}\ \bibnamefont
  {Fedorenko}}, \bibinfo {author} {\bibfnamefont {Y.~N.}\ \bibnamefont
  {Luponosov}}, \bibinfo {author} {\bibfnamefont {A.~V.}\ \bibnamefont
  {Bakirov}}, \bibinfo {author} {\bibfnamefont {M.~A.}\ \bibnamefont
  {Shcherbina}}, \bibinfo {author} {\bibfnamefont {S.~N.}\ \bibnamefont
  {Chvalun}}, \bibinfo {author} {\bibfnamefont {A.~Y.}\ \bibnamefont
  {Sosorev}}, \bibinfo {author} {\bibfnamefont {L.}~\bibnamefont {Grodd}},
  \emph {et~al.},\ }\bibfield  {title} {\bibinfo {title} {Large-size
  single-crystal oligothiophene-based monolayers for field-effect
  transistors},\ }\href@noop {} {\bibfield  {journal} {\bibinfo  {journal} {ACS
  appl. Mater. Interfaces}\ }\textbf {\bibinfo {volume} {11}},\ \bibinfo
  {pages} {6315} (\bibinfo {year} {2019})}\BibitemShut {NoStop}%
\bibitem [{\citenamefont {Vialetto}\ \emph {et~al.}(2021)\citenamefont
  {Vialetto}, \citenamefont {Rudiuk}, \citenamefont {Morel},\ and\
  \citenamefont {Baigl}}]{Vialetto2021}%
  \BibitemOpen
  \bibfield  {author} {\bibinfo {author} {\bibfnamefont {J.}~\bibnamefont
  {Vialetto}}, \bibinfo {author} {\bibfnamefont {S.}~\bibnamefont {Rudiuk}},
  \bibinfo {author} {\bibfnamefont {M.}~\bibnamefont {Morel}},\ and\ \bibinfo
  {author} {\bibfnamefont {D.}~\bibnamefont {Baigl}},\ }\bibfield  {title}
  {\bibinfo {title} {Photothermally reconfigurable colloidal crystals at a
  fluid interface, a generic approach for optically tunable lattice
  properties},\ }\href@noop {} {\bibfield  {journal} {\bibinfo  {journal} {J.
  Am. Chem. Soc.}\ }\textbf {\bibinfo {volume} {143}},\ \bibinfo {pages}
  {11535} (\bibinfo {year} {2021})}\BibitemShut {NoStop}%
\bibitem [{\citenamefont {Ke}\ \emph {et~al.}(2023)\citenamefont {Ke},
  \citenamefont {Peng}, \citenamefont {Wu}, \citenamefont {Ren}, \citenamefont
  {Zhao}, \citenamefont {Sheng},\ and\ \citenamefont {Li}}]{Ke2023}%
  \BibitemOpen
  \bibfield  {author} {\bibinfo {author} {\bibfnamefont {S.}~\bibnamefont
  {Ke}}, \bibinfo {author} {\bibfnamefont {B.}~\bibnamefont {Peng}}, \bibinfo
  {author} {\bibfnamefont {R.}~\bibnamefont {Wu}}, \bibinfo {author}
  {\bibfnamefont {J.}~\bibnamefont {Ren}}, \bibinfo {author} {\bibfnamefont
  {Y.}~\bibnamefont {Zhao}}, \bibinfo {author} {\bibfnamefont {Q.}~\bibnamefont
  {Sheng}},\ and\ \bibinfo {author} {\bibfnamefont {H.}~\bibnamefont {Li}},\
  }\bibfield  {title} {\bibinfo {title} {Simulation of crystal nuclei at the
  liquid-air interface toward morphology control via surface tension},\
  }\href@noop {} {\bibfield  {journal} {\bibinfo  {journal} {J. Phys. Chem. C}\
  }\textbf {\bibinfo {volume} {127}},\ \bibinfo {pages} {17231} (\bibinfo
  {year} {2023})}\BibitemShut {NoStop}%
\bibitem [{\citenamefont {Hirvonen}\ \emph {et~al.}(2016)\citenamefont
  {Hirvonen}, \citenamefont {Ervasti}, \citenamefont {Fan}, \citenamefont
  {Jalalvand}, \citenamefont {Seymour}, \citenamefont {Vaez~Allaei},
  \citenamefont {Provatas}, \citenamefont {Harju}, \citenamefont {Elder},\ and\
  \citenamefont {Ala-Nissila}}]{Hirvonen2016}%
  \BibitemOpen
  \bibfield  {author} {\bibinfo {author} {\bibfnamefont {P.}~\bibnamefont
  {Hirvonen}}, \bibinfo {author} {\bibfnamefont {M.~M.}\ \bibnamefont
  {Ervasti}}, \bibinfo {author} {\bibfnamefont {Z.}~\bibnamefont {Fan}},
  \bibinfo {author} {\bibfnamefont {M.}~\bibnamefont {Jalalvand}}, \bibinfo
  {author} {\bibfnamefont {M.}~\bibnamefont {Seymour}}, \bibinfo {author}
  {\bibfnamefont {S.~M.}\ \bibnamefont {Vaez~Allaei}}, \bibinfo {author}
  {\bibfnamefont {N.}~\bibnamefont {Provatas}}, \bibinfo {author}
  {\bibfnamefont {A.}~\bibnamefont {Harju}}, \bibinfo {author} {\bibfnamefont
  {K.~R.}\ \bibnamefont {Elder}},\ and\ \bibinfo {author} {\bibfnamefont
  {T.}~\bibnamefont {Ala-Nissila}},\ }\bibfield  {title} {\bibinfo {title}
  {Multiscale modeling of polycrystalline graphene: A comparison of structure
  and defect energies of realistic samples from phase field crystal models},\
  }\href {https://doi.org/10.1103/PhysRevB.94.035414} {\bibfield  {journal}
  {\bibinfo  {journal} {Phys. Rev. B}\ }\textbf {\bibinfo {volume} {94}},\
  \bibinfo {pages} {035414} (\bibinfo {year} {2016})}\BibitemShut {NoStop}%
\bibitem [{\citenamefont {Taha}\ \emph {et~al.}(2017)\citenamefont {Taha},
  \citenamefont {Mkhonta}, \citenamefont {Elder},\ and\ \citenamefont
  {Huang}}]{Taha2017}%
  \BibitemOpen
  \bibfield  {author} {\bibinfo {author} {\bibfnamefont {D.}~\bibnamefont
  {Taha}}, \bibinfo {author} {\bibfnamefont {S.~K.}\ \bibnamefont {Mkhonta}},
  \bibinfo {author} {\bibfnamefont {K.~R.}\ \bibnamefont {Elder}},\ and\
  \bibinfo {author} {\bibfnamefont {Z.-F.}\ \bibnamefont {Huang}},\ }\bibfield
  {title} {\bibinfo {title} {Grain boundary structures and collective dynamics
  of inversion domains in binary two-dimensional materials},\ }\href
  {https://doi.org/10.1103/PhysRevLett.118.255501} {\bibfield  {journal}
  {\bibinfo  {journal} {Phys. Rev. Lett.}\ }\textbf {\bibinfo {volume} {118}},\
  \bibinfo {pages} {255501} (\bibinfo {year} {2017})}\BibitemShut {NoStop}%
\bibitem [{\citenamefont {Waters}\ and\ \citenamefont
  {Huang}(2022)}]{Waters2022}%
  \BibitemOpen
  \bibfield  {author} {\bibinfo {author} {\bibfnamefont {B.}~\bibnamefont
  {Waters}}\ and\ \bibinfo {author} {\bibfnamefont {Z.-F.}\ \bibnamefont
  {Huang}},\ }\bibfield  {title} {\bibinfo {title} {Grain rotation and coupled
  grain boundary motion in two-dimensional binary hexagonal materials},\
  }\href@noop {} {\bibfield  {journal} {\bibinfo  {journal} {Acta Mater.}\
  }\textbf {\bibinfo {volume} {225}},\ \bibinfo {pages} {117583} (\bibinfo
  {year} {2022})}\BibitemShut {NoStop}%
\bibitem [{\citenamefont {Harutyunyan}(2012)}]{Harutyunyan2012}%
  \BibitemOpen
  \bibfield  {author} {\bibinfo {author} {\bibfnamefont {A.~R.}\ \bibnamefont
  {Harutyunyan}},\ }\bibfield  {title} {\bibinfo {title} {Uniform hexagonal
  graphene film growth on liquid copper surface: Challenges still remain},\
  }\href@noop {} {\bibfield  {journal} {\bibinfo  {journal} {Proc. Natl. Acad.
  Sci. U.S.A.}\ }\textbf {\bibinfo {volume} {109}},\ \bibinfo {pages} {E2099}
  (\bibinfo {year} {2012})}\BibitemShut {NoStop}%
\bibitem [{\citenamefont {Granato}\ \emph {et~al.}(2022)\citenamefont
  {Granato}, \citenamefont {Greb}, \citenamefont {Elder}, \citenamefont
  {Ying},\ and\ \citenamefont {Ala-Nissila}}]{Granato2022}%
  \BibitemOpen
  \bibfield  {author} {\bibinfo {author} {\bibfnamefont {E.}~\bibnamefont
  {Granato}}, \bibinfo {author} {\bibfnamefont {M.}~\bibnamefont {Greb}},
  \bibinfo {author} {\bibfnamefont {K.~R.}\ \bibnamefont {Elder}}, \bibinfo
  {author} {\bibfnamefont {S.~C.}\ \bibnamefont {Ying}},\ and\ \bibinfo
  {author} {\bibfnamefont {T.}~\bibnamefont {Ala-Nissila}},\ }\bibfield
  {title} {\bibinfo {title} {Dynamic scaling of out-of-plane fluctuations in
  freestanding graphene},\ }\href
  {https://doi.org/10.1103/PhysRevB.105.L201409} {\bibfield  {journal}
  {\bibinfo  {journal} {Phys. Rev. B}\ }\textbf {\bibinfo {volume} {105}},\
  \bibinfo {pages} {L201409} (\bibinfo {year} {2022})}\BibitemShut {NoStop}%
\bibitem [{\citenamefont {Taha}\ \emph {et~al.}(2019)\citenamefont {Taha},
  \citenamefont {Dlamini}, \citenamefont {Mkhonta}, \citenamefont {Elder},\
  and\ \citenamefont {Huang}}]{Taha2019}%
  \BibitemOpen
  \bibfield  {author} {\bibinfo {author} {\bibfnamefont {D.}~\bibnamefont
  {Taha}}, \bibinfo {author} {\bibfnamefont {S.~R.}\ \bibnamefont {Dlamini}},
  \bibinfo {author} {\bibfnamefont {S.~K.}\ \bibnamefont {Mkhonta}}, \bibinfo
  {author} {\bibfnamefont {K.~R.}\ \bibnamefont {Elder}},\ and\ \bibinfo
  {author} {\bibfnamefont {Z.-F.}\ \bibnamefont {Huang}},\ }\bibfield  {title}
  {\bibinfo {title} {Phase ordering, transformation, and grain growth of
  two-dimensional binary colloidal crystals: A phase field crystal modeling},\
  }\href@noop {} {\bibfield  {journal} {\bibinfo  {journal} {Phys. Rev.
  Mater.}\ }\textbf {\bibinfo {volume} {3}},\ \bibinfo {pages} {095603}
  (\bibinfo {year} {2019})}\BibitemShut {NoStop}%
\bibitem [{\citenamefont {Cook}(1970)}]{Cook70}%
  \BibitemOpen
  \bibfield  {author} {\bibinfo {author} {\bibfnamefont {H.~E.}\ \bibnamefont
  {Cook}},\ }\bibfield  {title} {\bibinfo {title} {Brownian motion in spinodal
  decomposition},\ }\href@noop {} {\bibfield  {journal} {\bibinfo  {journal}
  {Acta Metall.}\ }\textbf {\bibinfo {volume} {18}},\ \bibinfo {pages} {297}
  (\bibinfo {year} {1970})}\BibitemShut {NoStop}%
\bibitem [{\citenamefont {Provatas}\ and\ \citenamefont
  {Elder}(2010)}]{Nik2010}%
  \BibitemOpen
  \bibfield  {author} {\bibinfo {author} {\bibfnamefont {N.}~\bibnamefont
  {Provatas}}\ and\ \bibinfo {author} {\bibfnamefont {K.}~\bibnamefont
  {Elder}},\ }\href@noop {} {\emph {\bibinfo {title} {Phase-Field Methods in
  Materials Science and Engineering}}}\ (\bibinfo  {publisher}
  {\href{https://onlinelibrary.wiley.com/doi/book/10.1002/9783527631520}{Wiley-VCH
  Verlag GmbH}},\ \bibinfo {year} {2010})\BibitemShut {NoStop}%
\end{thebibliography}%

\end{document}